\documentclass[12pt]{iopart}

\usepackage{graphicx}

\usepackage{color,soul}
\definecolor{red}{rgb}{1,0.,0.} 
\setulcolor{red}

\usepackage{amssymb}

\def\IR{\relax{\rm I\kern-.18 em R}}

\def\L{{\cal L}}
\def\F{{\cal F}}
\def\E{{\cal E}}
\def\G{{\cal G}}
\def\J{{\cal J}}
\def\A{{\cal A}}

\def\S{{\cal S}}
\def\<#1>{\langle #1\rangle}

\def\tobepublished#1{\vspace{28pt plus 10pt minus 18pt}
     \noindent{\small\rm To be published: {\it #1}\par}}

\font\frak=eufm10 scaled\magstep1
\def\goth #1{\hbox{{\frak #1}}}

\def\bdelta{{\mbox{\boldmath $ \delta$}}}
\def\bomega{{\mbox{\boldmath $ \omega$}}}
\def\bOmega{{\mbox{\boldmath $ \Omega$}}}
\newcommand{\da}{\hbox{$\Box$}\,}

\begin{document}

\title[Geometric treatment of electromagnetic phenomena]{Geometric treatment of electromagnetic phenomena in conducting materials: variational
principles}

\author{A Bad\'{\i}a\,--\,Maj\'os$^1$, J F Cari\~{n}ena$^2$ and C L\'{o}pez$^3$}
\address{$^1$Departamento de F\'{\i}sica de la Materia Condensada--I.C.M.A., Universidad de Zaragoza, Spain}
\address{$^2$Departamento de F\'{\i}sica Te\'orica,
  Universidad de Zaragoza, Spain}
\address{$^3$Departamento de Matem\'aticas,
   Universidad de Alcal\'a de Henares, Spain}
\ead{anabadia@unizar.es, jfc@unizar.es, carlos.lopez@uah.es}

\date{\today}

\begin{abstract}

The dynamical equations of an electromagnetic field coupled with a
conducting material are studied. The properties of the interaction
are described by {a classical field theory with} tensorial material laws in space-time geometry.
{We show that} the main features of superconducting response emerge in a natural way within the
covariance, gauge invariance and variational formulation
requirements. {In particular, the Ginzburg-Landau theory follows straightforward from the London equations when fundamental symmetry properties are considered. Unconventional properties, such as the interaction of superconductors with electrostatic fields are naturally introduced in the geometric theory, at a phenomenological level. The BCS background is also suggested by macroscopic fingerprints of the internal symmetries.}

It is also shown that dissipative conducting behavior may be approximately
treated in a variational framework after breaking covariance for
adiabatic processes. Thus, nonconservative laws of interaction are
{formulated} by a purely spatial variational principle, in a
quasi-stationary time discretized evolution. This theory justifies a
class of nonfunctional phenomenological principles, introduced for
dealing with exotic conduction properties of matter [Phys. Rev.
Lett. 87, 127004 (2001)].

\end{abstract}

\tobepublished{\JPA}
\maketitle

\section{{Introduction}}

Classical electrodynamics has been a commonplace in several areas of mathematical physics. Thus, many of the physical problems linked to the applications of function theory or differential geometry are taken from electromagnetism. In particular, we recall the very elegant formulation of the electromagnetic (EM) field in space time geometry. The full set of Maxwell equations may be simply expressed as $d\F =0$ and $\delta{\F}={\J}$, for ${\F}$ (the electromagnetic field tensor) a closed 2-form defined on the 4-dimensional Minkowski space, ${\cal J}$ the current density 1-form, and $\delta$ the so-called codifferential operator \cite{flanders}.
The closedness condition of $\F$ ($d\F =0$) allows a local
representation in terms of a potential 1-form ${\cal A}$, introduced by
$\F = d\A$. Maxwell laws may then arise as the Euler--Lagrange
equations for an action functional in terms of ${\cal A}$. Within
this geometric formulation of electromagnetism, the consideration of symmetry properties simplifies
and illuminates the theory. In particular, covariance (which
assures invariance of the equations under transformations of the
Lorentz group) is very convenient and gauge invariance (invariance
under transformations of the potential ${\cal A}\mapsto {\cal
A}+d\chi$) must be satisfied. 

All the above ideas are mainly established in the study of the electromagnetic (EM)
field in vacuum, but can also be of great help in the research of interactions with electric charges within macroscopic media. Thus, it is known that the most
relevant aspects of superconductivity: expulsion of magnetic fields,
zero resistance, flux quantization and the phase-voltage
relationship at the gap between superconductors, may be straightforwardly obtained
from gauge invariance considerations \cite{weinberg}.

Recently, and motivated by some experimental puzzles in superconducting physics, several additional tools, conventionally restricted to the research of EM fields and sources in vacuum, have been applied for the description of material laws. To be specific, space time covariance of the phenomenological equations of superconductivity has been considered as the appropriate framework for explaining the still unclear interaction of these materials with electrostatic fields. Essentially, and inspired in the principle of Lorentz covariance, when the electric and magnetic fields are treated at the same level, one predicts both electrostatic and magnetostatic field expulsion with a common penetration depth $\lambda$ \cite{govaertsst,govaerts,hirschprb}. This formulation has been used \cite{hirschprl} as a basis for explaining the so-called Tao Effect, an intriguing experimental observation, in which superconducting microparticles aggregate into balls in the presence of electrostatic fields \cite{taoprl}. However, both theoretical objections \cite{commentreply} as well as experimental results \cite{bertrand} raise concerns on the universality of the common $\lambda$ treatment.

We want to emphasize that the topic of electrostatic field expulsion was already addressed by the London brothers \cite{london35} in the early days of superconductivity. Nonetheless, owing to the lack of experimental confirmation \cite{london36}, they eventually decided to formulate their celebrated equations of superconductivity in a noncovariant form. Such a lack of relativistic covariance leads to theoretical difficulties, but they are conventionally avoided by postulating the absence of electrostatic charges and fields within the sample. This point of view has been adopted by the scientific community over decades, until the revived interest mentioned in the previous paragraph.

In the light of the above comments, it is apparent that finding a physically sound covariant expression for the material physical laws, in intrinsic geometric terms may be of help. In fact, it will be shown that this will not only release the theoretical dissatisfaction caused by using a somehow {\em amended theory}, but it will also allow to incorporate new physical phenomena in a natural way. In an effort to produce the most concise and general equations, we propose in this work, the use of differential forms on Minkowski space, within the area of electrical conductivity. However, it should be emphasized that relativistic exactitude is not meant to influence the kinetics of superconducting carriers. What one tries to do is to write the physics in the clearest form so as to get information on the underlying mechanism, through the macroscopic (phenomenological) electrodynamics. Being electrodynamics a fundamental interaction in the case of superconductors, one tries to scrutinize within the nature of the phenomenon, taking advantage of the EM field symmetries.

As a first approximation to the problem, we will restrict ourselves to linear effects in the material response, i.e. nonelectromagnetic degrees of freedom enter through the linear laws $\J=\Theta\cdot\F$ or $\J=\Xi\cdot\A +\omega$, with $\Theta\in{\cal T}_{1}^{\, 2}$ and $\Xi\in{\cal T}_{1}^{\, 1}$ field-independent tensors, and $\omega$ a gauge adjusting 1-form. Although simple, these {\em ansatzs} provide celebrated material laws.

The study of the law $\J=\Xi\cdot\A +\omega$, with the requirement of gauge independence, will straightforwardly lead to the phenomenological London \cite{london35} and Ginzburg-Landau (GL) equations in covariant form, for the trivial case $\Xi_{\,\mu}^{\,\nu}=\alpha\,\delta_{\,\mu}^{\,\nu}$ (here, $\delta_{\,\mu}^{\,\nu}$ stands for the Kronecker's symbol). This will serve
as a basis for postulating a Lagrangian density, still a
controversial topic related to the time dependent GL theory
\cite{zagrodzinski}. On the other hand, nonequal diagonal terms in $\Xi$ will suggest relativistic BCS effects, in the manner introduced in \cite{bertrand}. Unconventional degrees of freedom, such as electrostatic charges will be unveiled in our gauge independent proposal.

The study of the law $\J=\Theta\cdot\F$ will lead to the concept of dissipative interaction, and thus to the absence of a direct variational formulation. However, we show that under quasistationary evolution, and following prescribed covariance breaking, one may issue a 3D variational principle, in terms of differential forms, at least for certain nonconservative interactions. This concept will also allow us to address some open questions in superconductivity. In particular, we justify the use of restricted variational principles in applied superconductivity for the so-called hard materials. At a first approximation, these superconductors are treated by a nonfunctional law, in which the electrical resistivity jumps from zero to infinity, when a critical value in the current density is reached \cite{PRL}.

The work is organized as follows. First, some mathematical background material, regarding notation and operations with differential forms is recalled in section \ref{secmathback}. The presentation is conceived so as to provide the minimal tools for using the powerful geometric language in what follows. Also, the variational formulation of pure or coupled electromagnetic problems
is reviewed. For further application, we emphasize the relation between
gauge invariance and the structure of the Lagrangian density.
Section \ref{secmaterial} is devoted to the proposal of covariant and
gauge invariant material laws for conducting media [$\J(\F)$ and $\J(\A)$]. We show that a
variational formulation for the covariant Ohm's law does not exist within the electromagnetic sector, and that, on the contrary, superconducting dynamics finds a very appropriate host in such formalism. Then, in section \ref{secnoncovariant} we give the rules for breaking
covariance in the formulation from the geometrical point of view. We introduce a restricted variational theory and exploit the benefit
of using it by applying these ideas to exotic superconducting systems, in which the material law is a nonfunctional relation. The implications of our analysis and further proposals for the theoretical
studies on conducting materials are summarized in
section \ref{secdiscuss}.

Rationalized Lorentz-Heaviside units will be used for the electromagnetic quantities along this article (with $c=1$).

\section{Mathematical background}
\label{secmathback}
\subsection{Covariant formulation of electromagnetism}
\label{seccovariant} Covariance is a well established requirement
for fundamental physical theories. In the more general statement it
means that the physical laws can be formulated in intrinsic geometric terms,
and therefore they are independent of local coordinate descriptions,
the reference system. There are sometimes further requirements
associated to some Lie group invariance, e.g., the Galilean group in
Classical Mechanics or the Lorentz group in special relativity.

\subsubsection{Basic notation and definitions}

The classical geometric treatment of the Maxwell equations is
formulated in a space-time ma\-ni\-fold $\mathcal{M}$ endowed with
a flat Lorentzian metric $g$ [here we choose the signature $(-,+,+,+)$]. Recall that
a  pseudo-Riemannian metric $g$ in a manifold $M$  provides us
with an isomorphism of the  $C^\infty(M)$-module of vector fields
in that of 1-forms, $\widehat g:{\goth{X}}(M)\to \bigwedge^1(M)$,
given by $\<\widehat
  g(X),Y>=g(X,Y)$, which allows us to transport the scalar product from
${\goth{X}}(M)$ to $\bigwedge^1(M)$. We  shall use the same
notation $g$ for such a product
\begin{equation}
g(\alpha_1,\alpha_2)=g(\widehat g^{\,-1}(\alpha_1),\widehat
g^{\,-1}(\alpha_2))\,, \qquad \alpha_1,\alpha_2\in \bigwedge\,^1(M).
\end{equation}
We also recall that given an $n$-dimensional pseudo-Riemannian
manifold  $M$, we define for any index $p\leq n$ a
$C^\infty(M)$-linear  map $star$ from the $C^\infty(M)$-module of
 $p$-forms into that of $(n-p)$-forms by means of
(see e.g. \cite{CP})
\begin{equation}
\alpha \wedge {\star \beta} = (-1)^sg(\alpha , \beta )\  {\rm Vol}
\qquad \forall \,\, \alpha, \beta \in \bigwedge \,^p (M)
\end{equation}
where $g(\alpha , \beta )$ is the scalar product in $\bigwedge
\,^p (M)$, i.e. if $\alpha $ and $\beta $ are decomposable
$p$-forms, $\alpha=\alpha_1\wedge \cdots\wedge\alpha_p$ and
$\beta= \beta_1\wedge \cdots\wedge\beta_p$, then
$g(\alpha,\beta)=\det (g(\alpha_i,\beta_j))$.
$\rm Vol$ is our choice of the volume form associated to the
metric; it is given by
\begin{equation}
{\rm Vol}=\pm\sqrt{(-1)^s\det g}\ dx^1\wedge\cdots \wedge dx^n\,,
\end{equation}
where $s$ denotes the signature, the number of negative squares
appearing in the quadratic form associated with $g$ when written
in its diagonal form.

In affine  coordinates for the particular case of Minkowskian
space $\mathcal{M}$ (for which $n=4$ and  $s=1$), $\{ x^{\mu}\mid
\mu = 0,1,2,3 \}$, the signature is 1 and for $\alpha=\alpha_\mu\,
dx^\mu$ and $\beta=\beta_\mu\, dx^\mu$ in $\bigwedge ^1
(\mathcal{M})$, we find (recall summation over repeated indices)
\begin{equation}
g(\alpha , \beta ) = g^{\mu \nu} \alpha _{\mu} \beta _{\nu} =
-\alpha _0 \beta _0 + \alpha _i \beta _i
\end{equation}
and we make the choice
\begin{equation}
{\rm Vol} = {\sqrt {-\det g }}\   dx^1 \wedge dx^2 \wedge dx^3
\wedge dx^0  =
 dx^1 \wedge dx^2 \wedge dx^3\wedge dx^0\,,
\end{equation}
so that
 in a local basis
\begin{eqnarray}
\star (dx^0) = - dx^1\wedge dx^2\wedge dx^3\,, \quad &&\star(dx^1) =
- dx^2\wedge dx^3\wedge dx^0\,,
\nonumber\\
\star (dx^2) =  dx^3 \wedge dx^0\wedge dx^1\,, \quad &&\star(dx^3) =
- dx^0\wedge dx^1\wedge dx^2\,,
\end{eqnarray}
and more generally
\begin{eqnarray}
\star \beta =&& - \beta _0 \, dx^1\wedge dx^2\wedge dx^3  - \beta _1
\, dx^2\wedge dx^3\wedge dx^0 + \beta _2 \, dx^3\wedge dx^0\wedge
dx^1 
\nonumber\\
&&- \beta _3 \, dx^0\wedge dx^1\wedge dx^2\,.
\end{eqnarray}

Similarly for 2- and 3-forms, the star operator is determined by
\begin{eqnarray}
\star (dx^0 \wedge dx^1 ) = - dx^2 \wedge dx^3\,, \quad \star
(dx^0 \wedge dx^2 ) =  dx^1 \wedge dx^3\,,
\nonumber\\
\star (dx^0\wedge dx^3 ) = - dx^1 \wedge dx^2\,,
\quad \star (dx^1 \wedge dx^2 ) =  dx^0 \wedge dx^3\,, 
\nonumber\\
\star (dx^1\wedge dx^3 ) =  - dx^0 \wedge dx^2\,,
\quad \star (dx^2 \wedge dx^3 ) =  dx^0 \wedge dx^1\,,
\end{eqnarray}
and
\begin{eqnarray}
\star (dx^1\wedge dx^2\wedge dx^3) = - dx^0\,,\quad
&&\star(dx^0\wedge dx^1\wedge dx^2) = - dx^3\,,
\nonumber\\
\star(dx^0\wedge dx^1\wedge dx^3) =  dx^2\,, \quad
&&\star(dx^0\wedge dx^2\wedge dx^3) = - dx^1\,.
\end{eqnarray}
By defining the pairing for scalars as $g(1,1) = 1$ we also find
\begin{equation}
\star (1) =  dx^1\wedge dx^2\wedge dx^3\wedge dx^0 \,, \quad \star
(dx^0 \wedge dx^1\wedge dx^2\wedge dx^3 ) = -1 \, .
\end{equation}
Notice that if $\omega\in \bigwedge \,^p (M)$, then  $(\star\,
\circ\, \star)\omega =(-1)^{p(n-p)+s}\,\omega$ and, in particular,
in the Minkowskian case for which $n=4,s=1$, $(\star \circ
\star)\omega=(-1)^{p+1}\omega$.

As usual, the exterior derivative of a $p$-form over an $n$-dimensional manifold is defined by
\begin{equation}
d\omega={\displaystyle\frac{\partial \omega_{i_{1}\dots i_{p}}}{\partial x^j}}\,
dx^{j}\wedge dx^{i_{1}}\wedge\cdots\wedge dx^{i_{p}}\, .
\end{equation} 
On the other hand, the codifferential operator is defined by $\delta = \star d
\star$, it maps $p$-forms into $(p-1)$-forms, and $d\circ \delta +
\delta \circ d$ is the Laplacian (or D'Alembertian) of the
manifold.

The space of the physical variables $\F$ for the EM Classical
Field Theory is the set of closed 2-forms in $\mathcal{M}$. Let us
introduce some geometric objects and manifolds appropriate for a
detailed  description of the EM theory.  {This will serve as a link between the geometric language and the more conventional (analytical coordinate dependent) statement of the problem}. It is well known that
2-forms in $\mathcal{M}$ are sections for the vector bundle
$\lambda : T^*{\mathcal{M}} \wedge T^*{\mathcal{M}} \to
{\mathcal{M}}$ of skew-symmetric $(0,2)$ tensors over
$\mathcal{M}$. For a given fibre bundle $\gamma : E \to M$, we
denote by $J^1\gamma$ the first jet bundle \cite{Sa} of $\gamma$,
the manifold of equivalence classes of sections $s: M \to E$ with
first degree contact at a fixed point $m \in M$. $J^1\gamma$ is
the natural geometric framework for a system of first order
partial derivative equations (PDE). Elements of $J^1\gamma$ are
denoted by $j^1_ms$, the equivalence class of all sections with
zero and first order partial derivatives at the point $m$ equal to
those of $s$. For a given section $s$, $j^1s$ denotes de first jet
lift, a section $j^1s : M \to J^1\gamma$ defined by $j^1s(m) =
j^1_ms$. Local coordinates in $E$ adapted to the projection $\gamma$,
$\{ x^{\mu}, y^a \}$ with $\{x^{\mu}\}$ coordinates in $M$,
determine associated local coordinates in $J^1\gamma$, $\{ x^{\mu},
y^a, z^a_{\mu} = \partial _{x^{\mu}} y^a\}$. Now, it is more
transparent that a system of first order PDE $H(x^{\mu}, y^a,
\partial _{x^{\mu}} y^a) = 0$ represents just a submanifold
$P\subset J^1\gamma$.

In particular, local adapted coordinates for $\lambda :
T^*{\mathcal{M}} \wedge T^*{\mathcal{M}} \to {\mathcal{M}}$, $\{
x^{\mu}, \F_{\mu \nu} \}$ with $\F (m) = \F_{\mu \nu}(m)
dx^{\mu}\wedge dx^{\nu}$, determine local coordinates for
$J^1\lambda$

\begin{equation}
\{x^{\mu}, \F_{\mu \nu}, \G _{\mu \nu {\rm ,} \sigma} = \partial
_{x^{\sigma}}\F_{\mu \nu} \}
\end{equation}
Closed 2-forms, i.e., sections $\F$ for $\lambda$ fulfilling $d\F
= 0$, are such that their first jet bundle lift $j^1\F:
\mathcal{M} \to J^1\lambda$  goes to zero through
skew-symmetrization. The skew-symmetrization in $J^1\lambda$ is a
natural map $sk_3: J^1\lambda \to (T^*{\mathcal{M}})^{\wedge 3}$
defined by $sk_3(j^1_m\F) = d\F(m)$. In local coordinates,

\begin{equation}
sk_3 \circ j^1\F (m) = \frac {1}{3} \left(\G _{\mu \nu {\rm ,}
\sigma} + \G _{\sigma \mu {\rm ,} \nu} + \G _{\nu \sigma {\rm ,}
\mu}\right)(m)dx^{\mu}\wedge dx^{\nu} \wedge dx^{\sigma} \,\, .
\end{equation}
Similarly, the codifferential $\delta$ determines a natural map
$\tau : J^1\lambda \to T^*{\mathcal{M}}$ given by $\tau (j^1_m\F)
= \delta \F (m)$. A local coordinate expression can be obtained
through the previously presented set of coordinate relations for
the star operator.

\subsubsection{Maxwell equations}

{Below, we show that the standard expressions of the Maxwell equations in terms of vector calculus operators are recovered from the previous formalism when a particular coordinate system is specified.} The set of Maxwell equations is a system of first order PDE in
the EM field $\F$. Then, they can be geometrically described as a
proper submanifold  $P \subset J^1\lambda$, or equivalently, as a
family of geometric tensorial equations in $J^1\lambda$  whose set
of solution points determines $P$. More precisely, Maxwell
equations are  just $d\F = 0$ and $\delta \F - \J = 0$, where $\J$
represents a the 4-current density, either prescribed or related
to $\F$ through some material law. If $\J$ is prescribed $\J :
{\mathcal{M}} \to T^*{\mathcal{M}}$, then $P$ can  be
alternatively defined as $P = sk_3^{-1}(0) \cap \tau ^{-1}({\rm
Im}\J)$. On the other hand, if there is some relation $\J (\F)$,
$\J : T^*{\mathcal{M}} \wedge T^*{\mathcal{M}} \to
T^*{\mathcal{M}}$ (not necessarily linear), then $P = \{ j^1_m\F |
sk_3(j^1_m\F) = 0 \,\, {\rm and} \,\, \tau (j^1_m\F) = \J
(\F(m))\}$. 

In local affine coordinates
\begin{equation}
\F = -E_i \, d x ^0 \wedge d x^i + \frac {1}{2} \epsilon _{ijk}
B^i\,\, d x ^j \wedge d x ^k
\end{equation}
determines the electric and magnetic vector components of $\F$,
obviously coordinate (i.e. reference frame) dependent. Here
$\epsilon _{ijk}$ is the totally skew-symmetric, Levi-Civita,
tensor. 

Both $d\F = 0$ and $\delta \F = \J$ represent the
covariant version of Maxwell equations. Let us obtain their coordinate dependent version.

The equation $d\F=0$ is given by
\begin{eqnarray}
d\F =&& -\partial _{x^j}E_i\,  dx^0 \wedge dx^i \wedge dx^j + \frac
{1}{2}\epsilon _{ijk}\,
\partial _{x^0} B^i\, dx^0 \wedge dx^j \wedge dx^k 
\nonumber\\
&&+ \frac {1}{2} \epsilon _{ijk}\,
\partial _{x^l} B^iº,
dx^j \wedge dx^k \wedge dx^l=0\,.
\end{eqnarray}
Using the vector differential operator $\nabla = (\partial _{x^1},
\partial _{x^2},\partial _{x^3})$,
$d\F=0$ becomes
\begin{eqnarray}
\label{eqmaxwellgeom}
\left\{
\begin{array}{rl}
&\nabla \cdot{\bf B} = 0 \cr
& \nabla \times {\bf E} + \partial _{x^0} {\bf B } = 0\,.
\end{array}
\right.
\end{eqnarray}
On the other hand, $\delta \F = \J$ is given by
\begin{eqnarray}
\label{eqampere}
\delta \F =&& -(\partial _{x^1}E_1 +
\partial _{x^2}E_2 + \partial _{x^3}E_3 ) dx^0 + (-\partial _{x^3}B_2
+ \partial _{x^2}B_3 - \partial _{x^0}E_1) dx^1 
\nonumber\\
&&+ (-\partial _{x^1}B_3 + \partial _{x^3}B_1 - \partial _{x^0}E_2)
dx^2 + (-\partial _{x^2}B_1 + \partial _{x^1}B_2 - \partial
_{x^0}E_3) dx^3 
\nonumber\\
&&= \J \equiv -\rho dx^0 + J_i dx^i
\end{eqnarray}
and, in vector analysis notation,
\begin{eqnarray}
\label{eqmaxwell}
\left\{
\begin{array}{rl}
&\nabla \times {\bf B} - \partial _{x^0} {\bf E } = {\bf J} \cr
&\nabla \cdot{\bf E} =\rho \, .
\end{array}
\right.
\end{eqnarray}
{Notice that the chosen metric tensor $(-,+,+,+)$ determines the 1-form
representation for $\J$, so that its corresponding 4-vector
field is $\widehat g (\J) = \rho\,
\partial _{x^0} + g^{ij}J_j
\partial _{x^i}$. On the other hand, recall that $\delta \J =
0$, the continuity equation ($\partial_{t}\rho + \nabla \cdot{\bf J}=0$ in standard notation), is a consistency requirement for
$\delta \F = \J$, easily obtained from $\delta ^{\,2} \alpha= (-1) ^{p+1}
\star d^{\,2} \star\alpha = 0$ for any p-form $\alpha\in \bigwedge ^p$}. 
{For further use, we give below the wave propagation expressions both in intrinsic terms and in standard notation. They arise by taking the exterior derivative in equation} (\ref{eqampere}), i.e.
\begin{equation}
d\delta\F = d\J \, .
\end{equation}
This admits the following coordinate form in terms of the D'Alembertian operator for our metric ($\da\equiv -\partial^{2}_{tt}+\nabla^{2}$) 
\begin{eqnarray}
\label{eqwaves}
\left\{
\begin{array}{rl}
&\da {\bf B} = -\nabla \times {\bf J} \cr
&\da{\bf E} =\partial_{t}{\bf J}+\nabla \rho \, .
\end{array}
\right.
\end{eqnarray}

Let us {now} review the geometrical notation and properties of the vector potential.
The closedness condition $d\F = 0$ can be (locally) {\it
integrated}, and disappears from the theory, by describing the EM
field as the exterior differential of a 1-form $\A$, the
4-potential 1-form, $\F = d\A$. The potentials are local sections
for the cotangent bundle $\pi : T^*{\mathcal{M}} \to
{\mathcal{M}}$, and $\F = d\A$ is determined by the composition of
the first jet lift and skew-symmetrization map, $\F = sk_2 \circ
j^1\A$, with $sk_2: J^1\pi \to T^*{\mathcal{M}} \wedge
T^*{\mathcal{M}}$. We have natural adapted coordinates $\{x^{\mu},
A_{\mu}, A_{\mu , \nu}\equiv
\partial _{x^{\nu}} A_{\mu} \}$ at $J^1\pi$, and $\{x^{\mu}, A_{\mu}, F_{\mu \nu} =
A_{\mu , \nu}- A_{\nu , \mu} \}$ at $T^*{\mathcal{M}} \wedge T^*{\mathcal{M}}$. Then,
the electric and magnetic vectors are represented through the potential by
\begin{eqnarray}
\left\{
\begin{array}{rl}
&{\bf E} = -\nabla \phi - \partial _{x^0} {\bf A}\,,\cr& {\bf B} = \nabla \times {\bf
A}
\end{array}\right.
\end{eqnarray}
with $\A \equiv- \phi\, dx^0 + A_i\, dx^i$. Note that in terms of our
chosen metric tensor $(-,+,+,+)$ the vector
field corresponding to the potential 1-form is $\widehat g (\A) = \phi \,\partial _{x^0} + g^{ij}\,A_j\, \partial _{x^i}$. 

On the other hand, the local representation of the EM field through
a local {\it integral} potential 1-form gives way to nonuniqueness,
with local gauge invariance symmetry ${\A}' = {\A} + d\chi$;
quotient by this gauge invariance determines the classical physical
degrees of freedom. Thus, it is important to notice that the potential
$\A$ is not directly a physical quantity at the classical level. However, recall that topological
obstructions ($\F$ being closed but not exact) associated to the
configuration of the problem have sometimes experimental
consequences; this is, for instance, the case of the well-know
Aharonov-Bohm effect, in which charged quantum particles are
influenced by the circulation of ${\bf A}$ in a multiply-connected
region where the electromagnetic field vanishes. From the physical point of view, gauge invariance is an additional requirement of
the theory whenever the potential appears, either in the material
laws (phenomenological or fundamental), interaction of the EM field
with conducting samples, or in the variational theories, where the
action functional must be gauge invariant.

\subsection{Variational principles of electromagnetism}
\label{secvariation}

{The existence of a variational formulation is well known for
fundamental theories and mandatory if one wishes to connect the classical and quantum levels.} In many cases, additional degrees of freedom
interacting with the EM field will also have their own kinetic and
potential Lagrangian terms, and the  energy conservation law
reflects the possible transfer between the EM and other energies. However, it must be stressed that phenomenological theories do not always permit a variational
formulation, because dissipation of the EM energy can be balanced by
generation of other kind of energy (usually thermodynamical), and
the corresponding degrees of freedom may be disregarded in the
theory, i.e., the system under consideration is open. This section
is devoted to review some examples of
Lagrangian theories for the EM field: free, interacting with prescribed
4-current, and with an additional scalar field. A number of specific features will be outlined for  their application to the proposal of material laws in the following one.

\subsubsection{Free EM theory}

The Lagrangian function for the free Field Theory, $\L : J^1\pi
\to {\mathbb R}$, is given by 
\begin{equation}
{\L}\,{\rm Vol}= {\displaystyle\frac{1}{2}{\F}\wedge \star \F = \frac{1}{2}\left( E^{2}-B^{2}\right)\,{\rm Vol}}
\end{equation}
Notice that, being in fact a real
function in $T^*{\mathcal{M}} \wedge T^*{\mathcal{M}}$, $\L \circ
sk_2$ is constant along the fibres
of $sk_2: J^1\gamma \to T^*{\mathcal{M}} \wedge T^*{\mathcal{M}}$.
This property is nothing but the gauge invariance at an algebraic
level, $\L$ takes the same value for two jets $j_m^1\A_1$ and
$j^1_m\A_2$ whose difference $j^1_m \A_2 - j^1_m\A_1$ (the jet
bundle has a natural affine structure) is symmetric, i.e., $sk_2
(j^1_m \A_1) = sk_2(j^1_m \A_2)$. This also means that the
Lagrangian is singular, and given a solution section $\A$ for the
Euler-Lagrange equation, we can build physically equivalent
solutions by adding to $\A$ arbitrary sections $\A_0$ whose first
lift is in the kernel of $sk_2$. Obviously, $\A_0$ are nothing but
closed one forms. Maxwell equations are first order PDE in the
EM field $\F$, and therefore are second order PDE in the potential
$\A$. Euler-Lagrange equations for a first order Lagrangian are
second order. The second order jet bundle $J^2\gamma$ of a fibre
bundle $(E,\gamma, M)$ is defined in a similar way to $J^1\gamma$,
taking now equivalence classes of sections up to second order
derivatives. Now, second order PDE are in geometric terms
submanifolds of $J^2\gamma$. The geometric description of the
Euler--Lagrange equations for a classical field theory goes
through the Poincar\'e-Cartan form $\Theta _{\L} = d_{S_V}\L + \L
{\rm Vol}$ (see \cite{Sa} for a detailed description of the
geometric treatment of the Euler-Lagrange equations in classical
field theories), which is an $n$-form in $J^1\gamma$ for $n ={\rm dim}
M$, and $S_V$ is a vector valued $n$-form generalizing the
vertical endomorphism in tangent bundles, defined for each volume
form $V$ in $M$. From $\Theta _{\L}$, the Euler-Lagrange form,
$\E_{\L} = (\gamma _2^1)^* (d\L\wedge {\rm Vol}) + d_{h} \Theta _{\L}$
with $d_h$ the {\it total} derivative mapping $r$-forms in
$J^1\gamma$ into $(r+1)$-forms in $J^2\gamma$, happens to be an
$(n+1)$-form in $J^2\gamma$. The geometric equation $(j^2s)^*
[\E_{\L}] = 0$ for unknown $s$ represents the second order
partial differential equations (PDE) fulfilled by sections $s$
of $\gamma:E \to M$ making stationary the action functional $S
(s)= \int \L(j^1s) {\rm V}$, that is, they are the
Euler-Lagrange equations of the variational principle. In local
coordinates $\{x^{\mu}, y^a,z^a_{\mu},z^a_{\mu \nu}\}$ in $J^2\gamma$

\begin{equation}
\E _{\L} = \left( \frac {\partial \L}{\partial y^a} - \frac
{d}{dx^{\mu}} \frac {\partial \L}{\partial z^a_{\mu}}\right)
dy^a\wedge {\rm V} \qquad \frac {d}{dx^{\mu}} = \partial
_{x^{\mu}} + z^a_{\mu} \partial _{y^a} + z^a_{\mu \nu} \partial
_{z^a_{\nu}}
\end{equation}
and its components are the well known Euler--Lagrange equations in
coordinate form.

For the case of EM field theory, we find Gauss' and Amp\`ere's law in vacuum

\begin{equation}
\partial _{x^0} \left(\frac {\partial \L}{\partial (\partial _{x^0}
\phi)}\right) + \partial _{x^j} \left(\frac {\partial \L}{\partial (\partial _{x^j}
\phi)}\right) \equiv  \partial _{x^j} \left(\partial _{x^i} \phi + \partial _{x^0}
A_j\right) = 0
\end{equation}
and

\begin{eqnarray}
\partial _{x^0} \left(\frac {\partial \L}{\partial (\partial _{x^0}
A_i)}\right) + \partial _{x^j} \left(\frac {\partial \L}{\partial (\partial _{x^j}
A_i)}\right) \equiv&&  \partial _{x^0} \left(\partial _{x^i} \phi + \partial _{x^0}
A_i\right) 
\nonumber\\
&&+ \delta ^{lj}\epsilon_{ilk}\epsilon^{kmn}  \partial _{x^j}\left(\partial _{x^m}A_n \right)
= 0
\end{eqnarray}

On the other hand, a direct application of Noether's theorem,
connected with the in\-va\-rian\-ce of the action under the Lorentz
group, leads to the concept of canonical energy-momentum tensor
\begin{equation}
\label{eqndeftensor}
\Theta^{\mu}_{\nu}=\frac{\partial{\L}}{\partial(\partial_{\mu}A_{\rho})}
\frac{\partial A_{\rho}}{\partial x^{\nu}}
-{\delta}^{\mu}_{\nu}{\L} \, .
\end{equation}
As usual, a symmetrized and gauge invariant version is preferred, in
order  to ease interpretation. Thus, we will use the field
symmetrizing technique \cite{belinfante}
\begin{equation}
T^{\mu\nu}\equiv \Theta^{\mu}_{\sigma}g^{\sigma\nu}+\partial_{\sigma}\left(
F^{\mu\sigma}A^{\nu}\right) \, ,
\end{equation}
from which the conservation law $\partial T^{\mu\nu}/\partial x^{\mu} = 0$ follows immediately.

The $00$ component of the (symmetrized) energy-momentum tensor
$T^{\mu \nu}$ \cite{Sa} is $T^{00} = \frac {1}{2} ({\bf E}^2 + {\bf
B}^2)$, the classical EM energy. In local coordinates, the zero
component of $\partial _{x^{\nu}} T^{\mu \nu}$ represents the
conservation of EM energy, and the spatial components are the
conservation of EM momentum. In particular, the continuity equation
\begin{equation}
\partial _{x^0} T^{00} + \nabla \cdot ({\bf E} \times {\bf B}) = 0
\end{equation}
is the balance between the energy density time variation and power
flow; integration of ${\bf E} \times {\bf B}$ along the boundary
of a compact region measures the power transfer out of it.
Obviously, this is not a conserved quantity for non
Lorentz-invariant Lagrangians.

\subsubsection{Electromagnetic field with external sources}

In the presence of sources (prescribed electrical charge and current
densities) for the EM field, an additional term ${\cal A}\wedge
\star {\J}$ in the Lagrangian, i.e.
\begin{equation}
{\L}\,{\rm Vol}= {\displaystyle \frac
{1}{2}{\F}\wedge \star \F - {\cal
A}\wedge \star {\J}}
\end{equation}
determines the
modified Euler--Lagrange equations
\begin{equation}
\delta \F =  \J\,.
\end{equation}
Notice that $\L {\rm Vol}= \frac {1}{2}{\F}\wedge \star \F - {\cal
A}\wedge \star {\J}$ is no longer constant along the fibres of the
skew-symmetric projection $sk_2$. Thus, for a gauge transformation
$\A \mapsto \A + d\chi$,

\begin{equation}
\L{\rm Vol} \mapsto \L{\rm Vol} + d\chi \wedge \star \J = \L{\rm Vol} + d(\chi \star \J) - \chi
d \star \J \, .
\end{equation}

The current density must fulfill the continuity equation $\delta \J
= 0$ in order to maintain the gauge invariance of the action
functional. Then, the Lagrangian is modified by a divergence term, which
does not affect the dynamics of the system. We stress that gauge
invariance is a fundamental ingredient of the theory, not only
forcing consistency for the Maxwell equations, but also
determining transformation properties for the Lagrangian densities
or the material laws.

In local coordinates the term $-{\cal A}\wedge \star {\J}$ takes the
form
\begin{equation}
\left(-\rho\, \phi + {\bf J}\cdot {\bf A}\right){\rm Vol}
\end{equation}
as it can be easily computed from the action of $\star$ presented
{above}.

In this case, the Euler--Lagrange equations take the form
\begin{equation}
\partial _{x^0} \left(\frac {\partial \L}{\partial (\partial _{x^0}
\phi)}\right) + \partial _{x^j} \left(\frac {\partial \L}{\partial
(\partial _{x^j} \phi)}\right) \equiv  \partial _{x^j}
\left(\partial _{x^j} \phi + \partial _{x^0} A_j\right) = \frac
{\partial \L}{\partial \phi}\equiv -\rho
\end{equation}
and
\begin{eqnarray}
\partial _{x^0} \left(\frac {\partial \L}{\partial (\partial _{x^0}
A_i)}\right) + \partial _{x^j} &&\left(\frac {\partial \L}{\partial
(\partial _{x^j} A_i)}\right)\equiv  \partial _{x^0}
\left(\partial _{x^i} \phi + \partial _{x^0} A_i\right)
\nonumber\\
 &&+ \delta
^{lj}\epsilon_{ilk}\epsilon^{kmn}  \partial _{x^j}\left(\partial _{x^m}A_n
\right)
 = \frac {\partial \L}{\partial A_i}\equiv  J_i\, ,
\end{eqnarray}
obviously equivalent to equation (\ref{eqmaxwell}).

Eventually, the energy balance equation becomes

\begin{equation}
\label{eqEnergyOpen}
\partial _{x^0} T^{00} + \nabla \cdot ({\bf E} \times {\bf B}) =  - {\bf
E}\cdot {\bf J} \, ,
\end{equation}
showing that there is a transfer between EM energy and other
modes. This may correspond to reversible storage of energy by
charged particles, irreversible thermodynamical losses, etc. ${\bf
E}\cdot {\bf J}$ is sometimes called thermodynamical activity.

\subsubsection{Coupling with a scalar field: the Klein-Gordon equation}

In some problems, the current density will not be prescribed, but arise as a
consequence of charged particles moving in the EM field according to
the electromagnetic force. New degrees of freedom have to be
incorporated to the Lagrangian density by writing $\J$ in terms of
the particle positions and velocities, and adding a kinetic term for
the masses of the particles and {possibly} a potential interaction term
between them. If, at a macroscopic level, the number of particles
allows to consider a continuum charge and current density, the total
system will be described by the EM and {\it fluid} fields. This
could be a good approximation for some physical systems such as low
density plasmas. Nevertheless, the usual interaction of EM with
matter is still unsatisfactorily described in this way. Charged particles
(electrons) move in a material lattice, with which they interact,
and interchange momentum and energy. Thus, new degrees of freedom
(vibrational, for instance) should be incorporated to the model.
This may cause serious difficulties, and in some instances a
phenomenological material law may be of great help.

Just as a starting point for our subsequent proposal (section
\ref{secGL}), we recall the simplest theory that couples the
electromagnetic phenomenon and a relativistic {\em material field}
$\psi$ {in covariant form.} Thus, if one considers a scalar spin-0 field
representation for the dynamics of the charged particles, the
basic Lagrangian \cite{itzykson} may be written as 
\begin{equation}
{\cal L}{\rm Vol}={\displaystyle \frac{1}{2}}{\F}\wedge \star \F
+{\displaystyle \frac{\hbar^2}{2m_{0}}}d\bar{\psi}\wedge \star\, d\psi
-{\displaystyle \frac{m_0}{2}}\bar{\psi}\psi\,{\rm Vol}-V(\bar{\psi}\psi){\rm Vol} +{\cal L}_{int}{\rm
Vol} \quad 
\end{equation}
with
\begin{equation}
{\cal L}_{int}{\rm Vol}\equiv -{\cal A}\wedge \star\,\left\{ {\displaystyle\frac{\hbar}{2m_{0}}}\left[ iq(\bar{\psi}\, d\psi - \psi\, d\bar{\psi})- {\displaystyle \frac{q^{2}}{\hbar}}\bar{\psi}\psi {\cal
A}\right]\right\}
\end{equation}
Note that ${\cal L}$ has been split up as  ${\cal L}_{EM}+{\cal
L}_{\psi}+{\cal L}_{int}$, indicating the self-interactions of
both fields, and a coupling term between them. The potential term
$V(\bar{\psi}\psi)$ may also incorporate reversible interactions of
the particle field with other degrees of freedom, as the
underlying material lattice.

As one can easily check, the Euler--Lagrange equations of the
corresponding  action integral, when ${\cal A}$, $\psi$ and its complex conjugate $\bar{\psi}$ are
taken as independent variables, are nothing but the coupled
Maxwell and Klein-Gordon equations. Just, one has to identify the
electromagnetic current density (take $\partial_{\A}$ in the Lagrangian) with the KG current density times
the basic electric charge ($q$), plus a vector potential related
term, i.e.
\begin{equation}
{\J}={\displaystyle\frac{\hbar}{2m_{0}}}\left[i q(\bar{\psi}\,d\psi -\psi\, d\bar{\psi})- {\displaystyle \frac{2 q^{2}}{\hbar}}\bar{\psi}\psi {\cal
A}\right] \, .
\end{equation}
\section{Covariant material laws in conducting media}
\label{secmaterial}

Interaction of the EM field with matter is described by some material law, an equation
determining the response of the medium through the appearance of charge and current
densities under an applied EM field, usually generated by sources which can be
typically considered far away form the area of interest. In principle,
in order to propose simple material laws one must take care of preserving the basic
rules of any EM theory, that is, covariance and gauge invariance. As already discussed, the existence of a variational principle depends on the possibility to
treat additional degrees of freedom associated to other kinds of energy that balances
the possible dissipation (or creation) of EM energy. In general, this will not be the case for a
phenomenological theory in which we are exclusively interested in the EM sector, but
there should be a Lagrangian density for {more} fundamental theories that try to consider all
the degrees of freedom present in the system under study.

Below, we present a material law fulfilling explicitly the first two requirements,
covariance and gauge invariance, which could describe the classical response of conducting matter, but which does not allow a variational principle.
{Afterwords, we will introduce several laws allowing a variational formulation, which will immediately lead to the concept of superconductivity. The geometrical treatment will allow a natural upgrading of the theory, so as to infer the covariant GL equations, as well as a first indication of the BCS background.}

\subsection{Nonvariational tensorial laws $\J (\F)$}

In the following, we will consider an {\it external} EM field $\F
_e$ ge\-ne\-ra\-ted by some given sources outside a particular
region $Q$ of the space, which are incorporated to the problem
through appropriate boundary conditions on the boundary $\partial
Q$, and a sample material in $R_0 \subset Q$, generating an
additional EM field $\F _r$ as a response to the applied
excitation. Here $\partial Q$ is taken far away from $R_0$, so that
the EM material response $\F _r$ can be neglected there. The local
current density $\J$ within the sample, with compact support in $R_0$,
will be determined by some material law $\J (\F)$, $\F = \F _e +
\F _r$. Therefore, we consider that the physical system is not
isolated, being fed by the external sources through the
boundary, and possibly with dissipation in the sample, i.e.
transfer of EM energy into another kind (thermodynamical,
mechanical, etc.). In general, the Lagrangian density obtained by replacing
$\J$ in the Lagrangian used for prescribed
currents will not produce the correct Euler--Lagrangian
equations. In fact, widely used material laws determine Maxwell
equations which are not variational. Then, dissipative force densities
are added to the free Euler--Lagrange equations by just writing
the current density as $\J(\F)$ in equation (\ref{eqmaxwell}). The simplest choice is a
local first order approximation. In order to preserve the geometric
flavor, a quite general law (under the pointed restrictions) may
be handled as the tensorial equation, by introducing  a $(2,1)$
type tensor $\Theta$ contracted with the EM $(0,2)$ tensor $\F$ to
determine the 1-form current density $\J$:
\begin{equation}
i_{X}\delta\F=\<\F,i_{X}\Theta>\equiv i_X\J
\;\;\Leftrightarrow\;\; J_{\mu} = \Theta _{\mu}^{\nu \rho}\,
\F _{\nu \rho}
\end{equation}
Different components of the $\Theta$ tensor represent well known
EM versus matter interaction behaviors. Thus 
\begin{equation}
-\rho = \Theta _0^{0j}\, \F_{0j} + \Theta _0^{ij}\, \F_{ij}
\end{equation}
contains both electric and magnetic polarizability, while
\begin{equation}
J_i = \Theta _i^{0k} \F_{0k} + \Theta _i^{jk} \F_{jk}
\end{equation}
describes Ohm's law and magneto-conductivity. 

In passing, we  note
that the celebrated covariant form of Ohm's law \cite{zhang}
$J^{\mu}+u^{\mu}u^{\nu}J_{\nu}= ({1}/{R})\F^{\mu\nu}u_{\nu}$,
where $u^{\mu}$ is the fluid four-velocity, corresponds to the
previous equation when $u^{i}=0$, $u^0 =1$ and $\Theta _i^{0j}=(
{1}/{R})\, \delta_{\, i}^{\, j}$.

Eventually, the set of (Maxwell) equations to be solved in this obviously covariant  and gauge
invariant model of electromagnetic interaction are
\begin{equation}
d\F =
0\, ,\qquad i_{X}\delta \F = \<\F,i_{X}\Theta> \;\;\forall X\,.
\end{equation}
For instance, if one goes again
to the case of linear isotropic conducting media, {in the absence of electrostatic charges,} the wave diffusion equations $d\delta \F = d(\Theta \cdot \F )$ become
\begin{eqnarray}
\label{eqnEuclideanmetalwaves}
\left\{
\begin{array}{rl}
&{\nabla^{2}}{\bf E}={\displaystyle \frac
{1}{R}\frac{\partial{\bf E}}{\partial t}+
\frac{\partial^{2}{\bf E}}{\partial t^{2}}}
\cr\cr
&{\nabla^{2}}{\bf B}={\displaystyle \frac
{1}{R}\frac{\partial{\bf B}}{\partial t}+
\frac{\partial^{2}{\bf B}}{\partial t^{2}}} \, .

\end{array}\right.%
\end{eqnarray}
They are the well known equations describing the
penetration of electromagnetic fields in conducting media, and have
to be solved supplemented by boundary conditions for the fields.

\subsection{Variational scalar laws $\J (\A)$}
\label{secvarscalar}

Let us now consider a {\it less classical} material law, determined by a
covariant relation between the current density and the local
potential field, $\J (\A)$. It {is} apparent that an additional current $\omega$ will have to be
considered in order to preserve gauge invariance of $\J$. In principle, such additional current may look a {purely mathematical} artifact;
however, as we will see, widely accepted classical models of
superconductivity are obtained in this way, and the new current can
be understood as the classical shadow of a more fundamental quantum
dynamical superconductivity theory. This geometric approach is
therefore  a natural way to generate classical approximate models
for macroscopic quantum properties of EM interaction with
matter.

We will consider purely variational problems, so that EM energy
variations are balanced by the energy variation of the additional
field. At a first stage, we will analyze some simple choices of the additional
current to be incorporated to the interaction term, and will not
consider the origin of this new current, i.e., the underlying field and its
corresponding kinetic and self-interaction terms will be neglected in the Lagrangian
dynamics. Later on, the new field will be incorporated to the theory by
a minimal coupling prescription.

\subsubsection{The London model ($d\omega = 0$)}
\label{seclondon}
Let us consider first the following EM Lagrangian density, with a
very simple choice of interaction term,
\begin{equation}
\label{eqLagLondon}
\L = \frac {1}{2} \F \wedge \star \F - \frac {1}{2\alpha} \J
\wedge \star \J \, ,
\end{equation}
where $\J$ is a 1-form
\begin{equation}
\J = \alpha\, \A + \omega \,,
\end{equation}
with $\alpha$ a constant parameter, and $\omega$ a closed 1-form ($d\omega = 0$) associated to some field not discussed yet. The EM
Euler--Lagrange equations become
\begin{equation}
\delta \F =  (\alpha \A + \omega) =  \J \,,
\end{equation}
determining Maxwell equations for the proposed material law. In
local coordinates, with $\A = -\phi\, dx^0 + A_i \,dx^i$ and $\omega =
\omega _0\, dx^0 + \omega _i dx^i$, we have
\begin{eqnarray}
\left\{
\begin{array}{rl}
&\partial_{x^i} E_i =  ( \alpha \phi - \omega _0)\,, \cr & (\nabla
\times {\bf B})_i = \partial _{x^0} E_i  +  (\alpha A_i + \omega
_i)\,.
\end{array}\right.
\end{eqnarray}

In order to preserve gauge invariance for $\J$, when one considers the gauge transformation $\A\mapsto \A ' = \A + d\chi$, $\omega$ should transform according to
\begin{equation}
\omega\mapsto \omega ' = \omega -  \alpha\, d\chi\,.
\end{equation}
Under this rule, the Lagrangian is manifestly gauge invariant. 

\paragraph{Topological properties}

The
integral of $\J$ (spatial part) along a curve inside the sample, has two
components, coming from the {\it potential} $\alpha \A$ and the
new current $\omega$. Notice that $\alpha \A + \omega$ is equivalent
to $\alpha (\A + d\chi) + (\omega - \alpha d\chi)$. Then, the circulation of $\J$ contains two gauge invariant components. In the present model, the circulation of $\omega$ vanishes for trivial topologies because of
Stokes' theorem. 

{Note that, the new current being closed, there is a gauge fixing where $\omega$ locally vanishes, by an adequate selection of $\chi$.} {However, the $\omega -$independent theory, which may be identified with the basic London equations} \cite{london35} {is somehow unsatisfactory, because one loses physical information.} In particular, notice that, although locally vanishing, $\omega$ can however contain a {\em topological charge} for non trivial topologies, like a hole on a plane as configuration space (or an infinite cylinder in 3-D space), i.e.
\begin{equation}
d\omega=0\Rightarrow {\bomega}\simeq\nabla f \not\Rightarrow \oint \bomega\cdot d{\bf l} = 0\,.
\end{equation}

\paragraph{Continuity}

At this stage, the continuity equation, $\delta \J = 0$, is not a consequence of gauge invariance, because
we have not considered yet the full action functional. Here
$\delta \J = 0$ is a consistency condition for Maxwell
equations, and determines the relation $\alpha \delta \A + \delta
\omega = 0$ between the EM potential and the new current. {According to this, if one takes the $\omega -$independent formulation, the gauge fixing f}reedom is lost, and one should work within the {\em Lorenz} gauge condition $\delta \A =0$.

\paragraph{Wave equations}

By using the definition of $\F = d\A$ and applying the
exterior derivative to the material Maxwell equations, in order to
eliminate the new current, we get
\begin{equation}
d \delta \F = \alpha \F \,.
\end{equation}
The left hand side represents de D'Alembertian of the EM field
($\delta d\F= 0$), so that we have obtained a wave propagation
equation with sources inside the sample. In local coordinates, and after appropriate identifications of the parameter $\alpha$ and the flux expulsion length scale $\lambda$ one has
\begin{eqnarray}
\label{londonwaves}
\left\{
\begin{array}{rl}
&\nabla \, ^2 {\bf B} = {\displaystyle\frac {\bf B}{\lambda ^2} + \frac {\partial
^2{\bf B}}{\partial t^2}}\,\, ,
\cr\cr
&\nabla \, ^2 {\bf E} = {\displaystyle\frac {\bf E}{\lambda ^2} + \frac {\partial
^2{\bf E}}{\partial t^2}}
\end{array}\right.
\end{eqnarray}
{Recall that the gauge-independent wave equations are insensitive to the $1-$form $\omega$, that disappears by the closedness condition. As ${\bf E}$ and ${\bf B}$ are observable quantities, these equations are a test for the soundness of the model in which $\omega$ is closed}

Under quasi-stationary
ex\-pe\-ri\-men\-tal conditions, where
the wave propagation can be disregarded, we get the
celebrated London equation
\begin{equation}
\label{eqLondon}
\nabla \, ^2 {\bf B} = \frac {\bf B}{\lambda ^2}\,.
\end{equation}
Additionally
\begin{equation}
\label{eqLondonE}
\nabla \, ^2 {\bf E} = \frac {\bf E}{\lambda ^2}
\end{equation}
represents a penetration of electric field, not usually
considered, but mandatory for covariant considerations
\cite{govaertsst,govaerts,hirschprb}. 

Although $\omega$ has been eliminated in the current model, it becomes clear that it is an unavoidable
geometric ingredient to deduce London's equations from a
variational principle while maintaining the gauge invariance. {In the following section, we study the more general case in which $d\omega \neq 0$. A different physical scenario will arise.}

\subsubsection{The modified London model ($d\omega \neq 0$)}

A more general model is obtained by relaxing the closedness condition for $\omega$, i.e., we allow for $d\omega \neq 0$. Now, one can find non vanishing circulations for $\omega$ even for trivial
topologies. Solutions for the material Maxwell equations
\begin{equation}
\delta \F = \alpha \A + \omega = \J\,,
\end{equation}
plus the continuity equation
\begin{equation}
\delta (\alpha \A + \omega) = 0
\end{equation}
will determine particular distributions of usual EM and the full current
density.

\paragraph{Topology}

The contribution of $\omega$ to the circulation relates to
magnetic flux inside the sample (recall that $\oint {\bf A}\cdot d{\bf l}=\int\!\int{\bf B}\cdot d{\bf s}$). This represents the existence of  vortices (quantized magnetic flux) associated to the so-called type
II superconductors \cite{tinkham}. The spatial
distribution of $\omega$ into compact regions where $d\omega \neq
0$ and a surrounding space with $d\omega =0$ becomes a
simplified classical model for the existence of quantum vortices
in type II superconductors. 
The fact that circulations embracing the {\it vortex} regions do not vanish is of importance, as it carries information about the superconducting field related to $\omega$.

\paragraph{Wave equation}

The wave equation reads $d\delta \F = \alpha \F + d\omega$, i.e.
\begin{eqnarray}
\left\{
\begin{array}{rl}
&\nabla \, ^2 {\bf B} = {\displaystyle\frac {\bf B}{\lambda ^2} + \frac {\partial
^2{\bf B}}{\partial t^2} - \nabla \times {\bomega}}\,\, ,
\cr\cr
&\nabla \, ^2 {\bf E} = {\displaystyle\frac {\bf E}{\lambda ^2} + \frac {\partial
^2{\bf E}}{\partial t^2} - \nabla {\omega_{0}} + \frac {\partial
\bomega}{\partial t}}
\end{array}\right.
\end{eqnarray}
Recall that the static or quasi-static approximations are obtained by neglecting
either all the time derivatives or just the second order ones, in
the previous formulas for covariant superconductivity. It is important to notice that in any case,
both $\nabla \times {\bomega}$ and $\nabla {\omega_{0}}$ should be
maintained, generalizing previous proposals \cite{hirschprb}. {Also, recall that though $\omega$ has been introduced for mathematical consistency, its temporal and spatial components $(\omega_{0},\bomega)$ become observable charge and current densities. By the moment, they have to be considered as phenomenological quantities, but below they will acquire a (more fundamental) microscopic significance.}

\subsection{A covariant, gauge invariant and variational model of
superconductivity $[\J (\A , \psi)]$} 
\label{secGL}

Let us now look for a field such that its associated current density
$\omega$ fulfills the former requirements of material law for
 EM interaction with matter. As we have already shown in section
\ref{secmathback}, the complex Klein-Gordon  field $\psi$ represents a simple choice
for that purpose, because it fulfills covariance, gauge invariance
and variational formulation requirements.  The above introduced free parameter
$\alpha$ will be adjusted so as to identify the correct gauge
transformation rule, and the interpretation of $\bar{\psi}\psi$ as the
density of superconducting carriers will bring us to the famous
Ginzburg-Landau equation, which here is proposed in a covariant and
gauge invariant framework. Recall that the superconducting carriers
(Cooper pairs) are spin-0 combinations of electrons, and thus, the
KG equation seems to be a reasonable starting point for a covariant field
theory of superconductivity. Thus, identifying $\psi$ as a charged
KG field, the {\em London current density term} $\omega$ may be
expected to be
\begin{equation}
\omega={\displaystyle \frac{i\hbar q}{2m_{0}}}\left(
\bar{\psi}\, d\psi-\psi\, d\bar{\psi}
\right) \, .
\end{equation}
Outstandingly, upon gauge transformations, $\omega$ verifies the required law
\begin{equation}
\label{eqnomegarule}
\omega\mapsto \omega'=\omega+{\displaystyle \frac{1}{\lambda^2}}d\chi
\end{equation}
if one defines the rules
\begin{equation}
\label{eqnAPsirule}
{
\psi\quad\mapsto\quad {\rm e}^{-iq\chi /\hbar}\psi\quad ;
\quad \bar{\psi}\quad\mapsto\quad {\rm e}^{iq\chi /\hbar}\bar{\psi}
}
\end{equation}
and
\begin{equation}
{\displaystyle \frac{1}{\lambda^2}\equiv\frac{\bar{\psi}\psi q^2}{m_0}} = - \alpha \, .
\end{equation}
To this point, $m_{0}$ and $q$ are just an effective mass and charge for  the KG particles. {The transformation rules in equation} (\ref{eqnAPsirule}) {correspond to the internal $U(1)$ symmetry of the charged field.}

As it has been discussed elsewhere \cite{weinberg,govaerts}, the KG particles may be interpreted as mediating Higgs bosons, whose fingerprint in the theory is the mass term
$\A\wedge\star\A$ for the electromagnetic potential.
In conclusion, the generalization of the
covariant London Lagrangian should read
\begin{equation}
\label{GLlagrangian}
{\cal L}{\rm Vol}={\displaystyle \frac{1}{2} \F\wedge\star\,\F
+\frac{\hbar^2}{2m_{0}}}\bar{D}\bar{\psi}\wedge\star\,{D\psi}
-V(\bar{\psi}\psi){\rm Vol}-{\displaystyle\frac{m_0}{2}}\bar{\psi}\psi\,{\rm Vol}\, ,
\end{equation}
as it follows from a minimal coupling principle, applied to the field $\psi$. To be specific, we have used the covariant derivative
\begin{equation}
D\equiv d+i{\displaystyle\frac{q}{\hbar}}\A
\end{equation}

Now, taking variations in equation (\ref{GLlagrangian}) respective to the 1-form $\A$,  and to the
scalar fields $\psi$ and $\bar{\psi}$ produces the set of Euler--Lagrange equations
\begin{eqnarray}
&{\displaystyle\frac{\partial}{\partial x^\nu}\frac{\partial{\cal
L}}{\partial (\partial A_\mu/\partial x^\nu)} =
\frac{\partial{\cal L}}{\partial A_\mu}}&
\nonumber\\
&\Downarrow&
\nonumber\\
(\delta \F)_{\mu} = &J_\mu \equiv -{\displaystyle\frac{\bar{\psi}\psi q^2}{m_0}A_{\mu}+
\frac{i\hbar q}{2m_0}}
\left(\bar{\psi}\partial_{\mu}\psi-\psi\partial_{\mu}\bar{\psi}\right)&
\nonumber\\ \nonumber\\
&\cdots&
\nonumber\\ \nonumber\\
&{\displaystyle\frac{\partial}{\partial x^\mu}\frac{\partial{\cal
L}}{\partial (\partial \bar{\psi}/\partial x^\mu)} =
\frac{\partial{\cal L}}{\partial \bar{\psi}}}&
\nonumber\\
&\Updownarrow&
\nonumber\\
&{\displaystyle\frac{\partial}{\partial x^\mu}\frac{\partial{\cal
L}}{\partial (\partial \psi/\partial x^\mu)} = \frac{\partial{\cal
L}}{\partial \psi}}&
\nonumber\\
&\Downarrow&
\nonumber\\
&{\displaystyle\frac{\hbar}{2m_0}}\left[\left( \partial_{\mu}-i{\displaystyle\frac{q}{\hbar}}A_{\mu}\right)\right] \left[\left(
\partial^{\mu}+i{\displaystyle\frac{q}{\hbar}} A^{\mu}\right)\right]\psi= {\displaystyle \frac{\partial V}{\partial
\bar{\psi}}}+{\displaystyle\frac{m_0}{2}}{\psi}&
\nonumber\\
&\Updownarrow&
\nonumber\\
&{\displaystyle\frac{\hbar}{2m_0}}\left[\left( \partial_{\mu}+i{\displaystyle\frac{q}{\hbar}}A_{\mu}\right)\right] \left[\left(
\partial^{\mu}-i{\displaystyle\frac{q}{\hbar}}A^{\mu}\right)\right]\bar{\psi}= {\displaystyle \frac{\partial
V}{\partial{\psi}}}+{\displaystyle\frac{m_0}{2}}\bar\psi \, .&
\end{eqnarray}

When the self interaction model is chosen to be
\begin{equation}
V=\mu\bar{\psi}\psi + \frac {1}{2}\nu (\bar{\psi}\psi)^{2}
\end{equation}
one gets the covariant Ginzburg-Landau-Higgs equations of
superconductivity \cite{govaerts}. In geometric notation we get
\begin{equation}
d\F =0 \quad , \quad\delta \F = -\frac {1}{\lambda^2} \A + \omega
\end{equation}
for the Maxwell equations ($\J$ includes a term due to the redistribution of carriers density), and
\begin{equation}
{\displaystyle\frac{\hbar}{2m_0}}
{\rm Tr}[g(\bar{D} ,D )](\psi) = [\mu +  \nu (\bar{\psi} \psi )] {\psi}
+{\displaystyle\frac{m_0}{2}}\psi
\end{equation}
for the carriers wave function in superconducting state. 

\subsubsection{Energy-momentum}
Starting from equation (\ref{GLlagrangian}) and keeping in mind that ${\cal L}$ depends on the fields $A^{\mu},\psi,\bar{\psi}$, one may calculate the full energy-momentum tensor from the expression
\begin{equation}
\label{eqndeftensorGL}
\Theta^{\mu}_{\nu}=\frac{\partial{\cal L}}{\partial(\partial_{\mu}A_{\rho})}\frac{\partial A_{\rho}}{\partial x^{\nu}}
+\frac{\partial{\cal L}}{\partial(\partial_{\mu}\psi)}\frac{\partial \psi}{\partial x^{\nu}}
+\frac{\partial{\cal L}}{\partial(\partial_{\mu}\bar{\psi})}\frac{\partial\bar{\psi} }{\partial x^{\nu}}
-{\delta}^{\mu}_{\nu}{\cal L} \, ,
\end{equation}
which will account for the self-energy of the electromagnetic and KG fields, plus the interaction. The calculation results in the symmetrized form
\begin{eqnarray}
T^{00}=&&{\displaystyle \frac{E^{2}}{2}+\frac{B^{2}}{2}}+
{\displaystyle \frac{\hbar^2}{2m_0}}\left(\partial_{0}\bar{\psi}\partial_{0}{\psi}+
{\displaystyle\frac{q^{2}}{\hbar^2}A_{0}\bar{\psi}\psi}+\bar{D}_{i}\bar{\psi}D_{i}{\psi}\right)
\nonumber\\
&&+{\displaystyle\frac{m_0}{2}}\bar{\psi}\psi+V(\bar{\psi}\psi) \, .
\end{eqnarray}
This is nothing but the GL free energy with relativistic effects. Recall that the {nonrelativistic} conventional expressions have been augmented not only by the particles rest energy, but also by the electrostatic terms $E^{2}$ and 
\begin{equation}
\left|
{\displaystyle\frac{iq}{\hbar}}A_{0}\psi\right|^{2}
\approx \lambda^2 \rho^{2}/2 \, ,
\end{equation}
as one could obtain from equation (\ref{eqLagLondon}) in the London limit.

\subsubsection{Nonrelativistic limit}

A straightforward calculation allows to obtain the nonrelativistic limit of the GL Lagrangian. This will produce a Schr\"{o}dinger-like equation for obtaining the low frequency limit of the time-dependent Ginzburg-Landau (TDGL) theory.

As a starting point, we split up the wave function in the form
\begin{equation}
\psi({\bf x},t)\equiv\Phi({\bf x},t) e^{-im_{0}c^{2}t/\hbar}
\end{equation}
with $\Phi$ the nonrelativistic part of the wave function, for which the relations
\begin{equation}
\label{eqnonreapprox}
{\displaystyle
\left|
i\hbar\frac{\partial\Phi}{\partial t}\right| \ll m_{0}c^{2}|\Phi|
\quad , \quad
|qA_{0}\Phi|\ll m_{0}c^{2}|\Phi|}
\end{equation}
must hold. Notice that in the previous formulas $c$ is not normalized, so as to ease quantitative comparison. Equations (\ref{eqnonreapprox}) mean that, compared to the rest energy, the nonrelativistic energy may be neglected, and that the potential is flat enough so as to avoid spontaneous pair creation.

By starting with equation (\ref{GLlagrangian}), separating the $\mu=0$ and $\mu= 1,2,3$ components, and implementing the above relations, one obtains
\begin{equation}
\label{GLlagrangiannonrel}
{\cal L}_{nonrel}\simeq{\displaystyle -\frac{1}{4} F_{\mu\nu}F^{\mu\nu}}
-{\displaystyle\frac{i\hbar}{2}}\left(\Phi\partial_{0}\bar{\Phi}
-\bar{\Phi}\partial_{0}{\Phi}\right)
+{\displaystyle\frac{\hbar^2}{2m_0}}\bar{{\bf D}}\bar{\Phi}\cdot{\bf D}\Phi
-V(\Phi\bar{\Phi}) \, ,
\end{equation}
with $\bf D$ the spatial part of the covariant derivative.
We remark that this Lagrangian produces the TDGL equations for the nonrelativistic limit, as well as the correct limit for time-independent solutions, in which case one recovers the {\em conventional} GL free energy. This point may be easily checked, just by examining the Euler-Lagrange equations. When ones considers variations respect to $\bar{\Phi}$, the familiar Schr\"{o}dinger-like equation for $\Phi$ is obtained. 
\begin{equation}
i\hbar{\displaystyle \frac{\partial \Phi}{\partial t}}=
{\displaystyle \frac{\hbar^2}{2m_0}
\left(
{\nabla}+{\displaystyle\frac{iq}{\hbar}}{\bf A}
\right)^{2}\Phi}
+\mu\Phi+\nu |\Phi|^{2}\Phi
\end{equation}
Recall that the inclusion of the mass term for the relativistic KG field has been essential in order to reach the above result.

\subsection{{Tensorial material law $\J (\Xi\cdot\A )$}}

The wave equations (\ref{londonwaves}), or their quasi-stationary approximations (\ref{eqLondon}) and (\ref{eqLondonE}),
determine the penetration profiles for the electromagnetic field in a Type I
superconductor. As mentioned before, equation (\ref{eqLondon}) is the widely accepted London model for the penetration of the magnetic
field and (\ref{eqLondonE}) its electric counterpart, which has been disregarded for decades since the pioneering works of the London brothers \cite{london35,london36}, who concluded that it wasn't physically sound to consider a finite decay length for the electric field.

Following the geometric orientation of this paper, we state that maintaining the magnetic penetration depth and rejecting the electric one (making it zero, as an {\em ansatz}) is inconsistent with the idea of covariance. This has also been recalled in references \cite{govaertsst,govaerts,hirschprb,hirschprl}, where several theoretical proposals along the lines of our study are given, some of them also connected with the experimental reality \cite{hirschprb,hirschprl}. However, a new controversy has appeared in the literature
about the subject of an electric penetration depth in superconductors. There are
firmly grounded theoretical reasons, as well as new experiments (see \cite{bertrand}) which support that, though nonzero, the typical
penetration length of the electric field is negligible with regard to the typical
one for the magnetic field. Roughly speaking, non superconducting
charges in the material, which do not play a role in the equilibrium electric current transport,
must be taken into account when an electric field generates an electrostatic response
in the sample. The order of magnitude of the ${\bf E}-$penetration depth related to
the normal charge density distribution happens to be much smaller than its magnetostatic counterpart, and this explains the lack of evidence for the
electrostatic phenomenon, at least within the experimental conditions considered.

In our phenomenological approach to the subject, the former physical considerations can be incorporated to the
material law for the superconducting state in a covariant and gauge invariant way. As a
preliminary proposal, let us consider an a bit more general material law than the one
introduced in section \ref{secvarscalar}. To be specific, let us concentrate on a still linear, but tensorial relationship $\J (\A )$ of the form
\begin{equation}
\J = \Xi \cdot \A  + \Omega \,
\end {equation}
with the $(1,1)$ tensor $\Xi$ depending upon two phenomenological constants in the form $\Xi _{\,i}^{\,i} = \alpha$, $\Xi _ {\,0}^{\,0}= \alpha + \beta$ and vanishing non diagonal
components, in a suitable affine coordinate system (the rest frame for the sample). The intrinsic tensor $\Xi$ could be computed in arbitrary coordinate systems through the
standard transformation law for $(1,1)$ tensors. 

\paragraph{Gauge invariance}
Notice that an additional 1-form $\Omega$ has been added to the material law for gauge invariance
considerations. In particular, under a gauge transformation $\A \mapsto \A + d\chi$, the
corresponding gauge transformation for $\Omega$ becomes

\begin{equation}
\label{gaugeOmega}
\Omega \mapsto \Omega - \Xi \cdot d\chi \, .
\end{equation}

Within the above model, we have the Maxwell equations
\begin{equation}
d\F = 0 \qquad \delta \F = \J = \Xi\cdot\A + \Omega \, .
\end{equation}
As stated before, these should be completed with the continuity equation $\delta \J = \delta (\Xi\cdot\A ) + \delta\Omega = 0$, which gives information about the properties of the additional current $\Omega$. On the other hand, the simplifying hypothesis $d\omega = 0$ assumed in the first London
model (section \ref{seclondon}) cannot be translated into this model because
\begin{equation}
d\Omega = 0 \qquad {\rm and} \qquad  d(\Omega - \Xi \cdot d\chi) = 0
\end{equation}
will be, in general, inconsistent for an arbitrary gauge function $\chi$, i.e., one can have $d(\Xi \cdot d\chi) \neq 0$ for a nontrivial tensor $\Xi$.

\paragraph{Wave equations}
A straightforward computation allows to obtain the particular form of the wave equations (\ref{eqwaves}) for this case. One has
\begin{equation}
\da{\bf B} = \frac {\bf B}{\lambda ^2} - \nabla \times {\bf \bOmega}
\end{equation}
and
\begin{equation}
\da{\bf E} = \frac {\bf E}{\lambda ^2} - \frac {\nabla \phi}{\nu ^2} - \nabla
\Omega _0+{\displaystyle\frac{\partial\bOmega}{\partial t}}
\end{equation}
with the definitions $\alpha = - {1}/{\lambda ^2}$ and $\beta = - {1}/{\nu ^2}$.

We emphasize that, under a
static configuration (all the time derivatives are neglected), the former equations become

\begin{equation}
\nabla ^2 {\bf B} = \frac {\bf B}{\lambda ^2} - \nabla \times {\bf \bOmega}
\end{equation}
and

\begin{equation}
\nabla ^2 {\bf E} = \left(\frac {1}{\lambda ^2} + \frac {1}{\nu ^2}\right){\bf E} - \nabla \Omega 
_0
\equiv \frac {1}{\lambda_e ^2}{\bf E} - \nabla \Omega_0 \, .
\end{equation}

As proposed in \cite{bertrand}, the electrostatic penetration depth combines the effects of the magnetic one ($\lambda$) and of another phenomenological constant ($\nu$). The microscopic origin of such dichotomy has been discussed in that work within the BCS theory framework. At the level of this article, what we can state is that $\Omega$ is the macroscopic manifestation of internal degrees of freedom beyond the Ginzburg-Landau $U(1)$ gauge invariant model. Just note that the tensorial gauge transformation rule in equation (\ref{gaugeOmega}) does not allow to introduce a complex scalar field ensuring a gauge invariant theory in the manner of equations (\ref{eqnomegarule}) and (\ref{eqnAPsirule}). Further aspects of this problem will be discussed elsewhere.

\section{Noncovariant material laws in conducting media}
\label{secnoncovariant}

{In the previous section, we have exploited the concept of relativistic covariance for studying electromagnetic material laws in conducting media. Having settled the basis for the geometrical description in 4-dimensional Minkowski space, we are ready to discuss about appropriate restrictions to 3-dimensional Euclidean space. This will be done below. The motivation for this part is to justify the use of restricted variational principles in the study of quasistationary conduction problems.}

\subsection{Breaking space-time covariance: reference frame}
\label{secbreaking}

In this section, a particular reference (rest or laboratory) frame
will be chosen, allowing to decompose the EM 2-form $\F$ into its
electric and magnetic vector field components. From a geometric
point of view, a reference frame corresponds to the choice  of
local coordinates (preferably affine) in the space-time manifold
${\mathcal{M}}$. However, in order to maintain the freedom about
the spatial coordinates (spatial covariance), we will consider the
splitting ${\mathcal{M}} = {\mathbb R} \times Q$, where $Q$ is the
three dimensional Euclidean space, or an open submanifold with
boundary if the particular properties of the system make it
desirable. $\mathbb R$ represents the absolute time for the
laboratory frame, and we have both natural projections $\pi _1$
and $\pi _2$ of ${\mathbb R} \times Q$ over each factor. Also, by
fixing a time value $t$, there is a trivial morphism $j_{t} : Q
\to {\mathcal{M}}$, $j_{t}(q) = (t,q)$, allowing to pull-back
forms in ${\mathcal{M}}$ into forms in $Q$. By doing it for each
point of an interval $[0,T]$, we can define one-parameter families
of forms in $Q$. Free selection of spatial frames means that the
EM theory is now developed in the 3-D tensor analysis framework.
When necessary, we will also fix a gauge in order to simplify some
equations, but gauge invariance of the theory must be maintained
even after breaking covariance. Different components of the
space-time forms are obtained by pulling-back the original form
(getting the space-like component) or its contraction with $\Gamma
= \partial _{t}$ (time-like component), the natural vector field
on the $\pi _2$ fibres ${\mathbb R}$.

With this notation we can obtain the corresponding components of
the EM field, the potential and the current density, associated to
the reference frame. The Hodge operator in $Q$, with the Euclidean
metric $g_E=\delta_{ij}$ (Kronecker's delta), will be denoted by
$*$ in order to distinguish it from the space-time $\star$ Hodge operator
in ${\mathcal{M}}$, and the exterior differential in $Q$ will be
denoted by ${\bf d}$ instead of the  $d$ in ${\mathcal{M}}$.
Similarly, the $(Q,g_E)$-codifferential will be denoted by
$\bdelta$. 

Now, we can define the  vectorial and scalar fields in
$Q$
\begin{eqnarray}
j_t^*(\F ) = * {\bf B} (t)\,, \quad &&j_t^*(i_{\Gamma }\F ) = -{\bf E}
(t)\,,
\nonumber\\
j_t^*(\A ) = {\bf A}(t)\,, \quad &&j_t^*( i_{\Gamma }\A ) = -\phi
(t)\,,
\nonumber\\
j_t^*(\J ) = {\bf J}(t)\,, \quad &&j_t^*( i_{\Gamma }\J ) = -\rho
(t)\,,
\end{eqnarray}
with the obvious identification of magnetic and  electric vector
fields, vector and scalar potential fields, as well as electric
vector current and charge densities. Spatial $\bf r$-dependence of
the fields has been avoided in the previous definitions for
simplicity. Maxwell equations in the reference frame take the
geometric form
\begin{eqnarray}
dF = 0 \,\,  \longrightarrow \,\,&&\{  \bdelta {\bf B} = 0 \,
, \quad {\bf d}{\bf E} +
\partial _t (* {\bf B})= 0\} \,,
\nonumber\\
\delta \F = \J \,\, \longrightarrow \,\, &&\{ \bdelta {\bf E} =
\rho \, , \quad
* {\bf d} {\bf B} - \partial _t{\bf E} = {\bf J} \}\,.
\end{eqnarray}
Quasistatic limits, with some field constant in time can be considered in the previous equations.

If one neglects $\partial _t (* {\bf B})$, i.e., electromagnetic energy is only stored in electric form, one reaches the so-called EQS (ElectroQuasiStatic) regime, in which
\begin{eqnarray}
\bdelta {\bf B} = 0 \,
, \quad {\bf d}{\bf E} = 0
\nonumber\\
\bdelta {\bf E} =
\rho \, , \quad
* {\bf d} {\bf B} - \partial _t{\bf E} = {\bf J} \, .
\end{eqnarray}
On the contrary, the MQS (MagnetoQuasiStatic) approximation corresponds to neglecting $\partial _t{\bf E}$. Then
\begin{eqnarray}
\bdelta {\bf B} = 0 \,
, \quad {\bf d}{\bf E}+
\partial _t (* {\bf B}) = 0
\nonumber\\
\bdelta {\bf E} =
\rho \, , \quad
* {\bf d} {\bf B} = {\bf J} \, .
\end{eqnarray}

Apparently, the EQS and MQS regimes arise when some characteristic speed in the problem is small as compared to $c$. In the general case, when no speed is neglected, energy is alternatively stored either in electric or magnetic forms, and one has a propagating wave. In the next section, we will concentrate on systems for which the MQS limit is attained. Our scenario will be as follows: some initial magnetostatic configuration is perturbed, giving place to a transient process in which electric fields and possible charge densities appear. Then, the system is driven to a final  magnetostatic configuration, and remains there until perturbed again. Dissipation ${\bf J}\cdot {\bf E}$ can appear
in the transient process and, although small, it cannot be neglected in the study.

\subsection{Spatial variational principles in quasistationary processes: the law $\J (\F )$}
\label{secspatial}

As it was seen before (sections \ref{secvariation} and
\ref{secmaterial}), the genuine variational formulation of
electrodynamics is done in $\mathbb{R}^{4}$ and in terms of the potential
1-form $\A$. However, some physical systems are successfully
analyzed in fixed reference (laboratory) frames, while keeping
spatial covariance. For instance, this is trivially true when one
focuses on the static equilibrium configuration of conservative
systems. Then, minimization of energy produces equations determining
the fields ${\bf E}$ and ${\bf B}$. What we show below is that such
idea may be generalized  to the quasi-stationary evolution of
dissipative systems. Under certain conditions, dynamical equations
produced by spatial variational principles are justified.

In Classical Mechanics it is well known that an initially
conservative system which is {\it slowly} drifted by an additional
small non conservative force, linear in the velocity, admits an
approximate variational principle. This represents an adiabatic
evolution, in which the energy, although not conserved, varies
slowly according to the adiabatic parameter. More
specifically, let us consider the dynamical equation

\begin{equation}
m\frac {d^2x}{dt^2} + \partial _xV = - \lambda \frac {dx}{dt}
\end{equation}
with small parameter $\lambda$. By adding to the Lagrangian $L=
({1}/{2} m)({dx}/{dt})^2 - V(x)$ the so-called {\it Rayleigh
dissipation} function\cite{goldstein} $({1}/{2}) t \lambda
({dx}/{dt})^2$ we find the Euler--Lagrange equation

\begin{equation}
m\frac {d^2x}{dt^2} + \partial _xV = - \lambda \frac {dx}{dt} - t\lambda \frac
{d^2x}{dt^2} \, ,
\end{equation}
which differs from the correct one in a negligible term for
small time intervals, because both $t$ and $\lambda$ are
considered small. 

Within a time interval $[0,\Delta T]$ the increments
fulfill the equation

\begin{equation}
m \Delta (\frac {dx}{dt}) + \langle\partial _x V\rangle  \Delta T = - \lambda
\Delta x -  \displaystyle{\frac{1}{2}}\Delta T \Delta (\lambda \frac {dx}{dt}) \, .
\end{equation}
The adiabatic hypothesis can be reformulated by saying that the
dissipative force $\lambda{dx}/{dt}$ varies slowly along the evolution. In Classical Mechanics, this is often used for
conservative systems with periodic orbits in which an adiabatic
evolution generates small variations of the parameters on each
cycle, e.g., the Poincar\'e map. It can also be used to perform a
numerical integration through time discretization and minimization
of the approximated action functional at each step, allowing to
apply minimizing techniques for the integral, usually more
reliable that a direct numerical integration of the differential
equations. {Recall} that the Euler--Lagrange equations are just
stationarity conditions for the action integral, but in any
reasonable physical variational system, the solutions are local
minimizers for the action, although possibly not global ones.

The {above method can also} be performed for an EM system with slow
drift between stationary configurations through a transient
material response to small source variations. The procedure could
be developed for quite general systems but, in order to fix the
ideas and taking into account the particular application which
follows, we will consider an MQS approximation for a
(type II super-)conductor, with vanishing $\bf E$ and $\rho$ in
the initial and final configurations. The transient evolution will
generate a small electric field, with possible local charge
density production, that will be neglected. 

We denote by ${\bf B}_0$ and
${\bf J}_0$ the initial stationary values, obviously fulfilling
$\nabla \times {\bf B}_0 = {\bf J}_0$. ${\bf B}_0({\bf r})$ and
${\bf J}_0({\bf r})$ fields are known functions, and we do not
need to introduce a potential field for the stationary
configuration. A potential vector field $\bf A$ is chosen {so as} to
describe the transient evolution, while the scalar field $\phi$ is
ignored in this approximation and should be determined by the null
charge density prescription. This can be interpreted as the selection of the temporal (Weyl) gauge. Then, along the evolution {one has}
\begin{eqnarray}
{\bf B}(t,{\bf r}) = {\bf B}_0({\bf r}) + \Delta _t{\bf B} ({\bf
r}) \quad&&{\tt with}\quad\Delta _t{\bf B} ({\bf r}) = \Delta {\bf B} (t,{\bf r})
\nonumber\\
{\bf J}(t,{\bf r}) = {\bf J}_0({\bf r}) + \Delta _t{\bf J} ({\bf
r}) \quad&&{\tt with}\quad\Delta _t{\bf J} ({\bf r}) = \Delta {\bf J} (t,{\bf r})\, .
\end{eqnarray}
Additionally 

\begin{equation}
\Delta _t{\bf B} = \nabla \times {\bf A}, \qquad {\bf E} =
-\partial _t {\bf A}
\end{equation}
as the representation in terms of the potential. 

Faraday's law
is automatically verified (geometric equation) while Ampere's law
is the one to be (approximately) determined through a variational
principle. The system under consideration will fulfill the
following conditions:

1) the finite sample material occupies the region $R_0$, $R_0
\subset Q$ with $Q$ also finite but $\partial Q$ far away form $R_0$
so that the material EM response decays to zero in $\partial Q$, and
the monitored sources are out of $Q$ determining the feeding of
the system through boundary conditions of the magnetic field in
$\partial Q$. Such experimental conditions are those of a PDE control
type problem, with a vectorial distributed parameter ${\bf B}_s$
form the sources and  {\it control} dynamical equations, Maxwell
equations, determining the response of the system.

2) the adiabatic hypothesis determines slow variations of the
sources form an initial value ${\bf B}_{s0}$ to a final value
${\bf B}_{s1}$ in a time interval $[0,T]$, with corresponding slow
evolution ${\bf B}_0 \mapsto {\bf B}_1$ and  ${\bf J}_0 \mapsto
{\bf J}_1$. Correspondingly, the transient electric field ${\bf
E}(t)$ is also small. The final values are determined by the
variables to be used in the analysis $\Delta _t{\bf B}$ and
$\Delta _t{\bf J}$, with ${\bf B}_1 = {\bf B}_0 + \Delta _T{\bf
B}$, ${\bf J}_1 = {\bf J}_0 + \Delta _T{\bf J}$. We also neglect
the EM wave by considering {\it instantaneous} material response,
that is, with typical time response negligible with regard to the
typical time parameter of the control variable. Moreover, and
similarly to the previous mechanical example,  along the adiabatic
evolution the electric field, playing the role of the dissipative
term, can be considered constant, i.e., ${\bf E} = {\bf E}({\bf
r})$ in $[0,T]$ and $\nabla \times {\bf B} = {\bf J}$.

The Maxwell equations must be complemented with some material law,
that we consider in the form $\Delta {\bf J} = {\bf G}({\bf E})$.
With $\bf E$ being small along the time interval, we make the
hypothesis that we can properly approximate the material law by
linearizing it to $\Delta {\bf J} \approx K \cdot {\bf E}$, with $K$ a
$3\times 3$ matrix representing the Jacobian of $\bf G$ in the
origin. Inhomogeneous $K({\bf r})$ could be considered, but here we
will choose the case of $K$ constant within the sample and
vanishing outside.

{Inspired by the above mentioned} mechanical Lagrangian, we define an MQS
Lagrangian density for $({\mathbb R}\times T^*Q, \pi, {\mathbb
R}\times Q)$ in order to determine a variational field theory for
the system

\begin{equation}
\label{eqnLagDissA}
\L \equiv \frac {1}{2} | \nabla \times {\bf A}|^2 + \frac {1}{2} t
\partial _t {\bf A}^{\tt T} \cdot K \cdot \partial _t {\bf A}
\end{equation}
where the super-index $\tt T$ denotes the transpose. {The associated} Euler--Lagrange equations become

\begin{equation}
\nabla \times (\nabla \times {\bf A}) + \partial _t \left( t K .
\partial _t{\bf A}\right) = 0 \, ,
\end{equation}
and working on them by substitutions we get

\begin{equation}
\label{eqnAmp}
\nabla \times \Delta _t {\bf B} = K \cdot {\bf E}  + t  \partial _t (K
\cdot{\bf E}) \approx \Delta _t {\bf J} \, .
\end{equation}
Above $\partial _t{\bf E}$ is neglected by the adiabatic hypothesis. Notice that we have
obtained Ampere's law, while Faraday's law was already fulfilled
by the potential representation.

The next step is to perform the time integration along $[0,T]$ in order
to get a purely spatial principle. We have chosen
$T$ small for a better approximation when neglecting $t  \partial _t
(K\cdot{\bf E})$. This allows to perform an approximate integral
by just considering mean values according to the initial and final
values of the magnetic field and current density, as well as constant $\bf E$ within the interval.
Writing the Lagrangian in terms of these variables we find the
minimization principle

\begin{equation}
{\rm min} \int _Q {\rm vol}\int _0^T \left( \frac {1}{2} (\Delta
_t{\bf B})^2 + \frac {1}{2} t {\bf E}.\Delta _t {\bf J}\right) dt
\end{equation}
After integrating in time according to the above
prescriptions, and avoiding global numerical factors, we get
\begin{equation}
\label{eqspatvar}
{\rm min}\int _Q \left( (\Delta _T{\bf B})^2 + T <{\bf E}>\cdot\Delta
_T{\bf J}\right) {\rm vol}
\end{equation}
a {\em purely spatial principle}. 

For the sake of completeness and consistency, let us
check that, under the hypothesis considered, the spatial
variational principle reproduces the correct dynamics. {For further application, we will}
do that in an unconventional way, {rewriting} the spatial
Lagrangian in terms of the variable $\Delta _T{\bf B}$ by using
both Ampere's law and the linearized material law. Notice that,
contrary to the prescription of Faraday's law through the use of
the potential in the space-time variational problem, we are here
prescribing Ampere's law by using it in the substitutions of the
spatial Lagrangian. 

Starting from 
\begin{equation}
L = \frac {1}{2} [ (\Delta _T{\bf B})^2
+ T (\nabla \times \Delta _T{\bf B})^{\tt T} \cdot K^{-1} \cdot (\nabla \times
\Delta _T{\bf B})] \, ,
\end{equation}
and performing the Euler-Lagrange equations
for the field theory in the spatial variables, we obtain
\begin{equation}
\Delta _T{\bf B} = - T \nabla \times \left[K^{-1} \cdot (\nabla \times
\Delta _T{\bf B})\right]
\end{equation}
which, through $(\nabla \times \Delta _T{\bf B}) = \Delta _T{\bf
J}$ and $K^{-1} \cdot \Delta _T{\bf J} = {\bf E}$, becomes

\begin{equation}
\nabla \times {\bf E} + \frac {\Delta _T{\bf B}}{T} = 0
\end{equation}
that is, the discretized version of Faraday's law. 

{We emphasize that} the above property is mainly grounded in the time discretization,
where the vector potential ${\bf A}$ can be rewritten as $-T{\bf
E}$, an integration which cannot be directly fulfilled in
space-time, and in the fact that the scalar potential has
disappeared form the formulation. In fact, the reader can write the
spatial Lagrangian in terms of the vector potential $\bf A$, i.e.,
prescribing the time discretized Faraday's law, and obtain in a
more conventional way Ampere's law as the Euler--Lagrange
equations for this Lagrangian. {We have sketched this in} (\ref{eqnLagDissA}--\ref{eqnAmp})

For a more physical interpretation of the result in equation (\ref{eqspatvar}), one can
identify a magnetic inertial term and a dissipative term, which
are balanced in order minimize the addition of magnetic flux changes and entropy production \cite{PRL}.

\subsection{Application to hard superconductivity: variational statement for nonfunctional $\{ {\bf E},{\bf J}\}$ laws}

\begin{figure}[!]
\centerline{\includegraphics[width=2.75in]{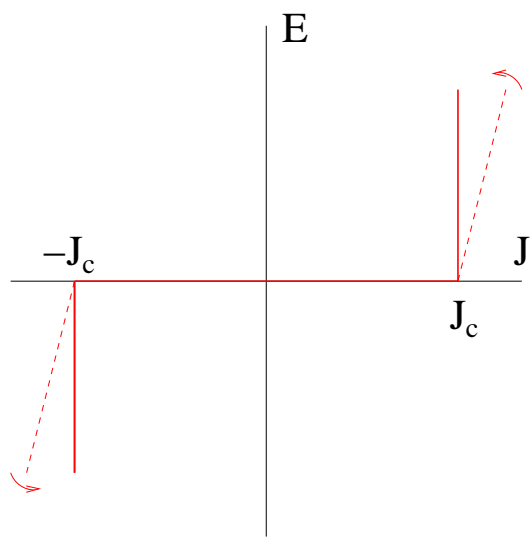}}
\caption{\label{fig1}$\{ E,J\}$ graph (conduction law) for a hard
type-II superconductor, according to Bean's model. {Vertical lines
correspond to infinite resistivity above $J_{c}$.}}
\end{figure}

Some physical systems are better described by general relations
(graphs) between their va\-ria\-bles, rather than by functional
ones. As a particular case of technological interest, we recall
the conduction property of the so-called hard type-II
superconductors. According to the phenomenological Bean's model
\cite{Bean}, the scalar components of ${\bf E}$ and ${\bf J}$,
when currents flow along a definite direction, are related by the
nonfunctional relation depicted in figure \ref{fig1}.
Notice that, if the electric field is nonzero at some point, one has
\begin{equation}
{\bf J}=J_{\rm c}{\displaystyle \frac{\bf E}{|{\bf E}|}}
\end{equation}
at such point. However, if $E = 0$ any value $J\in [-J_{\rm
c},J_{\rm c}]$ is allowed. {From the physical point of view}, the superconductor
reacts with a maximal current density flow to the application of
electric fields. When the excitation is canceled ($E\to 0$) the flow
may remain as a persistent nondissipative current. {Notice that the hard superconducting material displays the conventional quasistatic zero resistivity, until a certain level of current transport is demanded ($J_{c}$). Current densities above this threshold are no longer carried by }{\em supercharges}, {and a high electrical resistance is observed.}

A fundamental
justification of the {above model (so-called critical state model)}, the physical interpretation of the
material parameter $J_{\rm c}$ (critical current), and more
sophisticated versions may be found in \cite{PRL,bossavit,prigozhin} and references therein.

Here, we recall that variational methods are especially useful for
solving  and generalizing the above statement. To start with, we
interpret it as a more realistic limiting process (see figure \ref{fig1}).
Thus, the {\em hard material} allows lossless subcritical current
flow, while it reacts with a high resistivity $E=R (J-J_{\rm c})$
for $J>J_{\rm c}$ (analogously for negative values of $J$). The
harder the material, the higher value of $R$, and the more realistic
the graph law approximation. Then, one can start with equation (\ref{eqspatvar})
and notice that, as $R$ becomes larger, the second term also
increases with $J>J_{\rm c}$. In the limit of infinite slope,
this fact can be taken
into account by reformulating the variational principle as
\begin{figure}[b]
\centerline{
\includegraphics[width=5in]{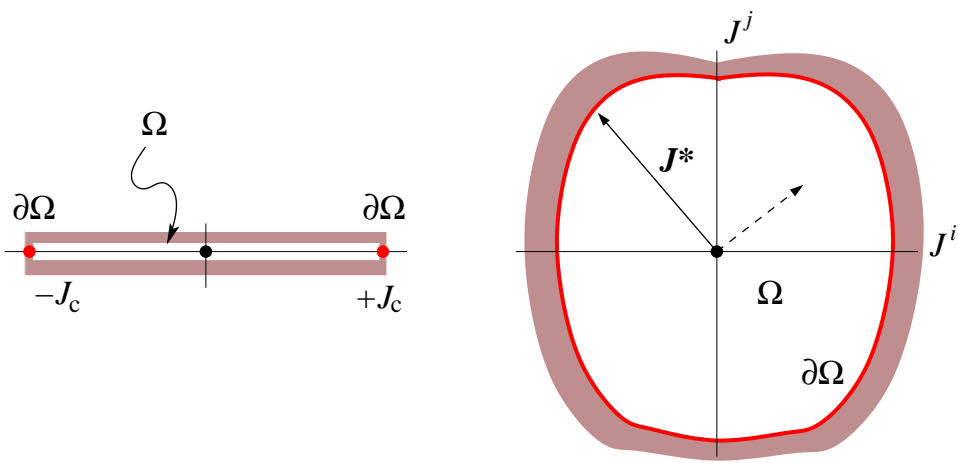}}
\caption{\label{fig2}Restriction sets for the current density
(${\bf J}\in\Omega$), corresponding to the behavior of hard
type-II superconductors. Optimal control solutions obey the
condition ${\bf J^{*}}\in\partial\Omega$.}
\end{figure}
\begin{eqnarray}
\label{eqvarsc}
&{\rm min}\quad \S = {\displaystyle \frac {1}{2}  \int _Q (\Delta {\bf B})^{2} {\,\rm vol}}&
\nonumber\\
\nonumber\\
&{\rm for \quad|\nabla \times {\bf B}|\leq J_c}&  \, .
\end{eqnarray}
The inequality (unilateral constraint) determines that the mathematical framework for the model is the so-called Optimal Control Theory \cite{Pon}, an
extension of the classical variational calculus for bounded
parameter regions. In the Optimal Control language, we have a
performance (cost) functional $\S$ to be minimized under the control
equation $\nabla \times {\bf B} = {\bf J}$ for bounded parameter
$|{\bf J}| \leq J_c$. As a very relevant property of this
variational interpretation, we remark that the control region for
the parameter may be understood as a physically meaningful
concept. Thus, one may pose the problem in the very general form
\begin{eqnarray}
\label{eqvarCS}
&{\rm min}\quad \S = {\displaystyle \frac {1}{2}  \int _Q  (\Delta {\bf B})^2 {\,\rm vol}}&
\nonumber\\
\nonumber\\
&{\rm for \quad\nabla \times {\bf B}\in \Omega}&  \, ,
\end{eqnarray}
with $\Omega\subset\mathbb{R}^{3}$ some restriction set, prescribed by the
underlying physical mechanisms. The conventional
statement, given by the graph in figure \ref{fig1}
is nothing but the particular case $\Omega = [-J_{\rm c},J_{\rm c}]$. This is depicted in
figure \ref{fig2}. {In the literature}, several possibilities for the set $\Omega$ have been studied, and
identified  as the fingerprint of dif\-fe\-rent physical
properties. For instance, elliptic restriction sets have been
shown to reproduce experimental observations related to
anisotropic current flow \cite{jap}, a rectangular set has been
used for producing the so-called double critical state model
\cite{Clem}, in which two independent critical current parameters
(parallel and normal) are used, etc.

\subsubsection*{Technical note}
The use of the principle (\ref{eqspatvar}) as a basis for obtaining (\ref{eqvarCS}) deserves some explanation. Not all the hypotheses used in the former case are straightforwardly translated. In particular, one has to recall that the linearization $\Delta{\bf J}=K\cdot{\bf E}$ has to be considered for excursions of $J$ around $J_c$. {However, according to figure} \ref{fig1}, when the electric field is reversed one has $\Delta{J}\simeq 2 J_c$. {Again, the} finite jump may be smoothed by taking a small time step. {Then}, the size of the region where $\Delta{J}= 2 J_c$ is negligible, and the fault has a very small weight in the integral to be minimized.

Finally, we recall that control equations
linear in the parameter, as the current case of interest, produce
the so-called {\em bang-bang solutions} \cite{Pon}, characterized
by the condition
\begin{equation}
{\bf J^{*}}\in\partial\Omega
\end{equation}
i.e. the optimal solutions take values at the boundary of the
allowed set.  For instance, the control variable jumps between $1$
and $-1$ when $\Omega=[-1,1]$ (see figure \ref{fig2}).

\begin{figure}[b]
\centerline{
\includegraphics[width=4in]{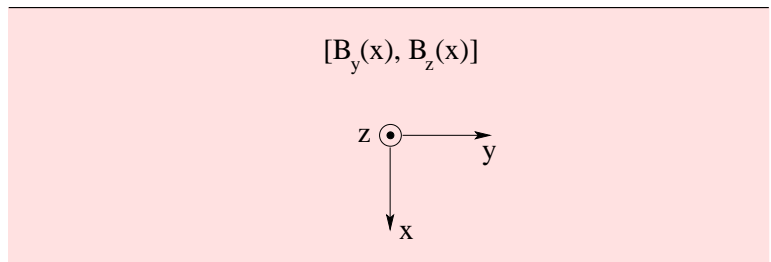}}
\caption{\label{fig3}Superconducting slab, subject to magnetic
field parallel to the surface.}
\end{figure}

\subsubsection*{Example: the infinite slab in parallel field.}

The optimal control approach to critical state problems in
superconductors has been sometimes misunderstood and qualified as a
more or less intuitive approach of restricted applicability. Among
other questions, it has been said that it lacks information, because
the electric field is absent of the theory. However, this quantity
is in the essence of the variational statement, which has been
obtained from the general Lagrangian, under the quasi-stationary
assumptions (section \ref{secspatial}). Below, we show with an example
that, indeed, the variational statement contains the electric field.
Thus, when the minimization process is performed, one obtains the condition that $\bf J$ belongs to the boundary, {and} also
a set of Lagrange multipliers (associated momenta in the Hamiltonian
formalism) closely related to the electric field. The comparison of our equations with a more conventional
approach that directly uses an ${\bf E}({\bf J})$ relation will
prove this aspect.

Let us consider the infinite superconducting slab depicted in  figure \ref{fig3}
($|x|<d/2; |y|,|z|<\infty$), and the problem of determining the electromagnetic
response to an excitation field parallel to the surface. Owing to the symmetry, one may
use the assumptions
\begin{eqnarray}
\left\{
\begin{array}{rl}
&{\bf B}=(0,B_{y}(x),B_{z}(x))\cr &{\bf J}=(0,J_{y}(x),J_{z}(x)) \, .
\end{array}\right.
\end{eqnarray}
Assuming isotropic conditions, equation (\ref{eqvarCS}) takes the form
\begin{eqnarray}
&{\rm min}\quad \S = {\displaystyle \frac {1}{2}  \int _Q  (\Delta {\bf B})^2 {\rm vol}}&
\nonumber\\
\nonumber\\
&{\rm for}\quad J=\;\;\left|{\displaystyle \frac{\partial {\bf
B}}{\partial x}}\right|\in \Omega=\left[0,J_c\right]&  \, .
\end{eqnarray}
Recall that the cost function depends on the field increment $\Delta {\bf B}={\bf B}-{\bf B}_{0}$ for each step of time, in the evolution of the system.

According to Pontryagin's maximum principle \cite{Pon}, this problem
is solved by combination of (i) the canonical equations for the
associated Hamiltonian
\begin{equation}
{\cal H}={\bf p}\cdot{\displaystyle \Delta{\bf J}}- {\displaystyle \frac
{1}{2}(\Delta {\bf B})^2} \, ,
\end{equation}
and (ii) {$\Delta {\bf J}^{*}={\bf J}^{*}-{\bf J}_{0}$ such that $H(\Delta {\bf J}^{*}) \geq H(\Delta {\bf J}) \; , \;
\forall\, {\bf J}\in \Omega$, i.e.,}
\begin{equation}
{\rm max}\left[{\bf p}\cdot{\displaystyle \Delta {\bf J}}\right]
\Rightarrow {\displaystyle \frac{\partial {\bf B}}{\partial
x}=J_{\rm c}\frac{\bf p}{p}} \, .
\end{equation}
This leads to the system
\begin{eqnarray}
\frac{\partial B_y}{\partial x}&=&J_{c}\frac{p_{y}}{p}
\nonumber\\
\frac{\partial B_z}{\partial x}&=&J_{c}\frac{p_{z}}{p}
\nonumber\\
\frac{\partial p_y}{\partial x}&=&\Delta B_y
\nonumber\\
\frac{\partial p_z}{\partial x}&=&\Delta B_z
\end{eqnarray}
By using the definition  $(p_y
,p_z)\equiv \Delta t (E_z ,-E_y)$ the system may be rewritten as
\begin{eqnarray}
\frac{\partial B_y}{\partial x}&=&J_{c}\frac{E_z}{E}
\nonumber\\
\frac{\partial B_z}{\partial x}&=&-J_{c}\frac{E_y}{E}
\nonumber\\
\frac{\partial E_z}{\partial x}&=&\frac{\Delta B_y}{\Delta t}
\nonumber\\
\frac{\partial E_y}{\partial x}&=&-\frac{\Delta B_z}{\Delta t}  \, .
\end{eqnarray}
Taking derivatives, and inserting the standard notation of dots and primes, one obtains
\begin{eqnarray}
\frac{\partial^{2} B_y}{\partial x \partial t}&=&J_{c}\frac{\dot{E_z}E-E_{z}\dot{E}}{E^{2}}
\nonumber\\
\frac{\partial^{2} B_z}{\partial x \partial t}&=&-J_{c}\frac{\dot{E_y}E-E_{y}\dot{E}}{E^{2}}
\end{eqnarray}
and
\begin{eqnarray}
 \frac{\partial^{2} B_y}{\partial x \partial t}&=&E_{z}''
\nonumber\\
\frac{\partial^{2} B_z}{\partial x \partial t}&=&-E_{y}''
\end{eqnarray}
and thus,
\begin{eqnarray}
\label{eq:secord}
E_{z}''&=&J_{c}\frac{\dot{E_z}E-E_{z}\dot{E}}{E^{2}}
\nonumber\\
E_{y}''&=&J_{c}\frac{\dot{E_y}E-E_{y}\dot{E}}{E^{2}}  \, .
\end{eqnarray}
Now, using polar coordinates in the plain,
\begin{eqnarray}
E_{y}=E\cos{\varphi}
\nonumber\\
E_{z}=E\sin{\varphi}
\end{eqnarray}
and taking space and time derivatives, it follows
\begin{eqnarray}
E_{z}''&=&E''\cos{\varphi}-2E'{\varphi}'\sin{\varphi}-E(\varphi ')^{2}\cos{\varphi}-E{\varphi}''\sin{\varphi}
\nonumber\\
E_{y}''&=&E''\sin{\varphi}+2E'{\varphi}'\cos{\varphi}-E(\varphi ')^{2}\sin{\varphi}+E{\varphi}''\cos{\varphi}
\nonumber\\
\dot{E}_{y}&=&\dot{E}\cos{\varphi}-E\dot{\varphi}\sin{\varphi}
\nonumber\\
\dot{E}_{z}&=&\dot{E}\sin{\varphi}+E\dot{\varphi}\cos{\varphi}
\, .
\end{eqnarray}
Eventually, back-substitution in equation (\ref{eq:secord}) leads to the system
\begin{eqnarray}
\left\{
\begin{array}{rl}
&E''=(\varphi ')^{2}E\cr
&2\varphi ' E'+ E\varphi ''=  J_{c}\dot{\varphi} \, .
\end{array}
\right.
\end{eqnarray}

These differential equations, together with suitable boundary
conditions, allow us to obtain the penetration profiles for the modulus
of the electric field  and for its angular direction. They have been
obtained from our variational approach, and fully coincide with the
expressions obtained by the ${\bf E}({\bf J})$ method in
reference \cite{brandt}.
\section{Conclusions and outlook}
\label{secdiscuss}

{In this article, we have shown that the geometric formulation framework of the classical electromagnetic field in terms of differential forms in Minkowski space may be extended to the study of conducting materials.}

{Related to recent intriguing experimental observations, and to the underlying dissatisfaction caused by a noncovariant theory for several aspects of superconducting electrodynamics, we have proposed a new phenomenological approach to the problem. We show that with the geometric jargon, the theory may be highly simplified. In the language of 1-forms, covariant superconductivity is merely a linear law which admits the inclusion of phenomenological constants and physical quantities. Such quantities allow a direct physical interpretation, as they are a part of wave equations in which they couple to the observable macroscopic fields ${\bf E}$ and ${\bf B}$.}

{Having clarified the basis of a covariant theory, we also present the complementary side of how covariance should be broken if required by the mathematical counterpart of some physical process. Thus, we show that quasistationary conduction problems may be treated in a spatial 3D covariant framework by pullback of the Minkowski space 1-forms to the $\mathbb R ^{3}$ Euclidean space. Taking advantage of this prescription, we have been able to justify the use of restricted variational principles in some problems of interest for applied superconductivity.}

Two different
possibilities for the material law have been considered, $\J (\F )$
and $\J (\A )$, i.e. the current density 1-form either depends on
the electromagnetic field 2-form or on the potential 1-form. {Being the simplest choice, linear dependencies have been considered.}

The {linear} law $\J (\F )$ is obviously covariant and gauge invariant by construction. However,
it is not variational. Only after breaking covariance, and under a quasi-stationary
approximation, one can issue a restricted variational principle for such a case. In
particular, we obtain an approximated spatial variational statement for linearly
dissipating systems (nonconservative forces are proportional to $\partial_{t}{\bf A}$,
whose temporal variation may be considered small).

The linear law $\J (\A )$ is covariant and allows a variational
statement  by endowing the electromagnetic field Lagrangian with
an interaction term of the form $\J\wedge\star\J$. However, {in this case,} gauge
invariance has to be required. By doing it, one is naturally led
to add new currents, generating the equations of superconductivity {($\J = \alpha\A + \omega$)}. The most relevant features of this
phenomenon are direct consequences of internal symmetries in the
field theory. Within the so-called London approximation, the main
field is $\A$. However, {the} 1-form $\omega$ is required by the
theory, coming from an additional field. Outstandingly, this field
has observable consequences (as the {presence of electrostatic charges and the} flux quantization condition),
and is sensible to the topology of the material. Just a step
further produces the so-called covariant Ginzburg-Landau theory.
If one identifies $\omega$ as a Klein-Gordon like probability
current density ($\omega \propto \bar{\psi}\, d\psi-\psi\, d\bar{\psi}$),
one has a conduction theory with two fields $(\A ,\psi)$, which
may be readily identified as the covariant and gauge invariant
generalization of the (non covariant) GL equations of
superconductivity. Here $\psi$ represents the wave function of the
superconducting carriers.

{In the first step, providing the simplest possible covariant expression for the conductivity of a material, we have proposed $\J = \alpha\A + \omega$. This fits many experimental observations, including the influence of electrostatic fields on superconductors} \cite{taoprl}. {However, accounting for other classical} \cite{london36} {and very recent experiments} \cite{bertrand}, {such proposal has been generalized to $\J = \Xi\,\cdot\A + \Omega$, with $\Xi$ a $(1,1)$ tensor. This form allows to unify the referred manifestations of superconductivity by either equal or nonequal phenomenological constants in the diagonal terms of $\Xi$. In this sense, when $\Xi$ is nontrivial, we argue that internal symmetries of the charge carriers, and gauge invariance are only compatible through a BCS approach. In this case, an additional field must be introduced, as not only the superconducting carriers are relevant. Nonsuperconducting charges may contribute to the static response of the material and their associated fields could be a matter of further research.}

Finally, we stress that the variational interpretation of {\em a priori nonvariational} conducting material
laws under adiabatic approximation has provided us with a method to treat {\em exotic} materials, in which the relation
between the fields is {well} determined through a graph. In particular, this has noticeable consequences in the phenomenological theory of
type-II superconductors. We have shown that the so-called Bean's model for hard type-II superconductors admits a variational formulation, grounded in basic properties of the electromagnetic Lagrangian. {This property is of utter importance in the field of applied superconductivity as it allows us to introduce numerical implementations for realistic systems, affected by finite size effects.} At the level developed in this work, the material properties are just included by augmenting the basic term $\F\wedge\star\F$ with a dissipation function contribution. Extensions of the theory in which the base Lagrangian includes the conservative terms of superconductivity ($\J\wedge\star\J ,\; \bar{\bf D}\bar{\Phi}\cdot{\bf D}{\Phi},\dots$) are expeditious.

Variational methods are shown to be equivalent to alternative
treatments of the problem, but offer a number of advantages. New mathematical tools, as
the optimal control theory, useful for discussing about the existence and form of the
electromagnetic problem solution, as well as for hosting numerical implementations, are
incorporated.

\section{Acknowledgments}

The authors acknowledge financial support from Spanish CICYT
(Projects BMF-2003-02532 and MAT2005-06279-C03-01).

\end{document}